\documentclass[aps,prx, twocolumn, floatfix,longbibliography,nofootinbib,superscriptaddress]{revtex4-1}
\usepackage[utf8]{inputenc}
\usepackage{natbib}
\usepackage{graphicx}
\usepackage{xcolor}
\usepackage[tbtags]{amsmath}
\usepackage[colorlinks, linkcolor=blue]{hyperref}
\hypersetup{colorlinks,allcolors=black}
\usepackage{amssymb}
\usepackage{amsmath}
\usepackage{gensymb}
\usepackage{float}
\usepackage{amsmath}
\usepackage{tabularx,graphicx}
\usepackage{epstopdf}
\usepackage{latexsym}
\usepackage{color, colortbl}
\usepackage{psfrag}
\usepackage{bbm}
\usepackage{bm}
\usepackage{titlesec}
\usepackage{dsfont}
\usepackage{feynmp}
\usepackage{slashed}
\usepackage{multirow}
\usepackage{mwe}
\usepackage[normalem]{ulem}

\newcommand{\bcen}{\begin{center}}
\newcommand{\ecen}{\end{center}}
\newcommand{\btab}{\begin{tabular}}
\newcommand{\etab}{\end{tabular}}
\newcommand{\bdes}{\begin{description}}
\newcommand{\edes}{\end{description}}

\newcommand{\beq}{\begin{equation}}
\newcommand{\eeq}{\end{equation}}
\newcommand{\bea}{\begin{eqnarray}}
\newcommand{\eea}{\end{eqnarray}}

\newcommand{\bary}{\begin{array}}
\newcommand{\eary}{\end{array}}
\newcommand{\benum}{\begin{enumerate}}
\newcommand{\eenum}{\end{enumerate}}
\newcommand{\bitem}{\begin{itemize}}
\newcommand{\eitem}{\end{itemize}}

\newcommand{\bfig}{\begin{figure}}
\newcommand{\efig}{\end{figure}}
\newcommand{\vev}[1]{{\left\langle #1 \right\rangle}}
\newcommand{\td}{\text{d}}

\newcommand{\mZ} {{\mathcal{Z}}}

\newcommand{\bra}[1]{{\langle #1 |}}
\newcommand{\ket}[1]{| #1 \rangle}

\newcommand{\eqn}[1] {eqn.~(\ref{#1})}

\newcommand{\sect}[1] {Section~\ref{#1}}
\newcommand{\Sect}[1] {Section~\ref{#1}}

\newcommand{\Fig}[1]{Fig.~\ref{#1}}

\makeatletter

\newcommand{\Rmnum}[1]{\expandafter\@slowromancap\romannumeral #1@}
\makeatother

\def\nd{{ \vphantom{\dagger}}}

\definecolor{darkred}{rgb}{0.4,0,0}

\makeatletter
\newsavebox{\@brx}
\newcommand{\llangle}[1][]{\savebox{\@brx}{\(\m@th{#1\langle}\)}%
  \mathopen{\copy\@brx\mkern2mu\kern-0.9\wd\@brx\usebox{\@brx}}}
\newcommand{\rrangle}[1][]{\savebox{\@brx}{\(\m@th{#1\rangle}\)}%
  \mathclose{\copy\@brx\mkern2mu\kern-0.9\wd\@brx\usebox{\@brx}}}
\makeatother

\begin{document}

\title{Spectral Form Factors of Topological Phases}
\author{Anurag Sarkar}
\email[E-mail:]{sarkara@iitk.ac.in}
\affiliation{Department of Physics, Indian Institute of Technology, Kanpur 208016, India}
\author{Subrata Pachhal}
\email[E-mail:]{pachhal@iitk.ac.in}
\affiliation{Department of Physics, Indian Institute of Technology, Kanpur 208016, India}
\author{Adhip Agarwala}
\email[E-mail:]{adhip@iitk.ac.in}
\affiliation{Department of Physics, Indian Institute of Technology, Kanpur 208016, India}
\author{Diptarka Das}
\email[E-mail:]{didas@iitk.ac.in}
\affiliation{Department of Physics, Indian Institute of Technology, Kanpur 208016, India}

\begin{abstract}

Signatures of dynamical quantum phase transitions and chaos can be found in the time evolution of generalized partition functions such as spectral form factors (SFF) and Loschmidt echos. While a lot of work has focused on the nature of such systems in a variety of strongly interacting quantum theories, in this work, we study their behavior in short-range entangled topological phases - particularly focusing on the role of symmetry protected topological zero modes. We show, using both analytical and numerical methods, how the existence of such zero modes in any representative system can mask the SFF with a large period (akin to generalized Rabi) oscillations, hiding any behavior arising from the bulk of the spectrum. Moreover, in a quenched disordered system, these zero modes fundamentally change the late-time universal behavior reflecting the chaotic signatures of the zero energy manifold.  Our study uncovers the rich physics underlying the interplay of chaotic signatures and topological characteristics in a quantum system.

\end{abstract}

\maketitle

\section{Introduction}

Ideas of thermalization and chaos have become pervasive in multiple sub-disciplines of physics, including classical and quantum many-body systems, quantum field theory, gravity and fluids\cite{Srednicki_PRE_1994, Rigol_AIP_2016, BORGONOVI20161, Das_PRL_2018,PollackPRL2020, Murugan_PRL_2021, polchinski2016, sss2, Dyer:2016pou, Nivedita_PRE_2020}. These questions become crucial to understand the phases of interacting quantum systems when they are either driven or are coupled to baths; however, methods of characterizing chaos are few and limited. The behavior of spectral form factor (SFF) in such systems has been particularly illuminating. Most interestingly, one finds that SFFs and its generalizations, such as Loschmidt echo \cite{sonner2017}, Fisher zeros \cite{2005, 2020} follow universal features independent of underlying microscopic details.  It is known that interacting many-body chaotic systems show a dip-linear~ramp structure in the SFF, which saturates at late times\cite{cotler2017, liu2018}. This behavior, however, has intriguing micro-structures which depend on the underlying symmetries, nature of interactions, dimensionality and localization properties \cite{sonner2017}. SFF has also been recently investigated even in non-interacting systems to decipher signatures of single particle chaos \cite{Winer:2020mdc, Liao:2020lac} and in field theories to determine the signatures of a critical point \cite{Nivedita_PRE_2020, Kudler-Flam2019JHEP}. This has led to interesting connections between a host of phenomena belonging to diverse physics ideas. However, the behavior of SFF vis-a-vis topological properties of a quantum Hamiltonian has been little explored. 

In this paper, we investigate the behavior of SFF in symmetry-protected topological systems. Here, phases are characterized by topological invariants, which are protected by discrete symmetries \cite{Ludwig_PS_2015, Chiu_RMP_2016, Hasan_RMP_2010, Qi_RMP_2011, asboth2016short}. A paradigmatic topological model is the Su-Schrieffer-Heeger (SSH) model \cite{ssh}, where the system hosts chiral symmetry-protected edge modes in the topological phase. We show that the existence of such zero-dimensional topologically protected eigenstates can fundamentally transform the SFF, exhibiting large time oscillations. We find, both numerically and analytically, that these are generalizations of Rabi oscillations, where the SFF locks between a few states of the complete many-body spectrum. We further show that this holds even in a higher-order topological insulator (HOTI) phases~\cite{benalcazar2017quantized}, where a $d$ dimensional topological phase hosts  $(d-2)$ dimensional boundary modes.

Given the quadratic nature of the Hamiltonians, our study also adds to the emerging area of understanding one-body chaos, recently explored in SYK-2\cite{Winer:2020mdc, Liao:2020lac} and in strongly coupled free gauge theories \cite{Chen:2022hbi}. In order to explore signatures of one-body chaos and its interplay with topological order, we study a variant of the SSH model where a subregion of the bulk is randomized with all-to-all hoppings while keeping the edge protected. The SFF, in this case, shows a dip followed by an early time oscillating exponential ramp reminiscent of one-body chaos in SYK-2 \cite{Winer:2020mdc}. This develops into an intermediate linear ramp, that plateaus at late times in both the trivial and topological phases. In the topological phase, while for any representative configuration, at late times there are Rabi oscillations; under ensemble averaging, these oscillations get destroyed given the random phase lags between various Rabi modes. This reflects the non-self-averaging characteristic of the SFF \cite{Prange} in the topological, yet disordered systems. Interestingly, this averaged asymptotic value in the topological phase is different from that in the trivial phase and depends only on the random matrix properties of the zero energy manifold. We end the paper with a perspective on how the interplay between topological features and chaotic signatures may be a rich playground to uncover a host of new phenomena in both lattice and field-theoretic quantum many-body systems. 

The paper is organized in the following way: in \Sect{SFF}, we discuss the general definition of spectral form factor (SFF) for a many-body chaotic system and for a free-fermionic eigenspectrum. In \Sect{model}, we study the feature of SFF in topological Hamiltonians, in particular with the presence of zero-dimensional boundary modes in the system. Then, we discuss how SFF behaves in a topological phase with a `chaotic' bulk-disordered background in \Sect{bulkssh}. We provide an intuitive understanding of our results by introducing an effective toy model and in parallel by random matrix theory results. We provide a summary of results and conclude in \sect{concl}. The stability of the disordered system and some detailed calculations has been provided in the Appendices \ref{apndxA} and \ref{apndxB}.

\section{Spectral Form Factor}\label{SFF}

\begin{figure}
\centering
 \includegraphics[width=0.85\linewidth]{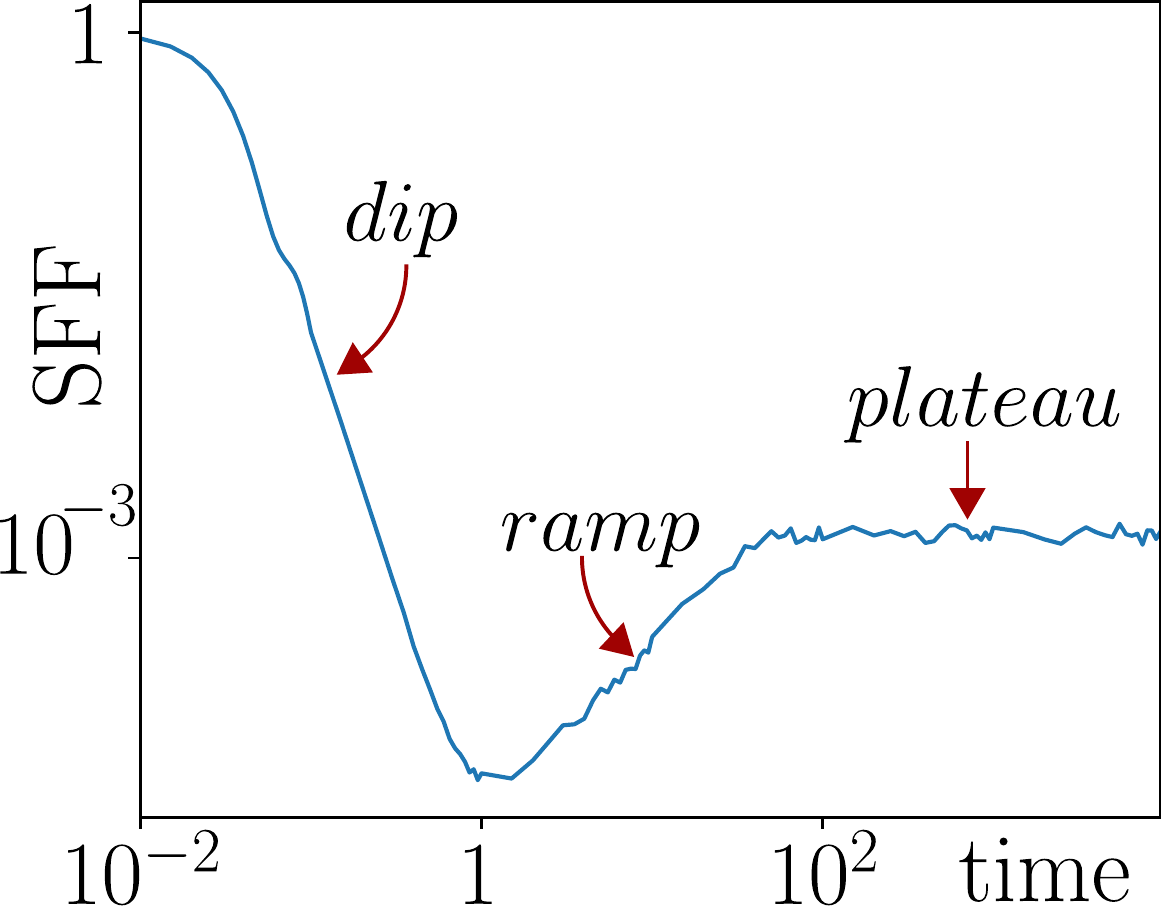}
	\caption{The ensemble-averaged SFF for Gaussian Unitary ensemble with $1000 \times 1000$ matrices, denoting dip, linear ramp and plateau. }
	\label{fig:sffGUE} 
\end{figure}

In general, the spectral form factor (SFF) of a system is defined as 
\beq
\text{SFF}(\beta, t) \equiv \mZ_2 (\beta, t) = \frac{Z(\beta+ i t)Z(\beta - i t)}{Z(\beta)^2},
\label{sffdef}
\eeq
where $\beta, t$ are the inverse temperature and real-time, respectively. $Z(\beta+ i t)$ denotes the generalized partition function. In terms of the many-body energy eigenstates $E_n$, we can then write:
\beq
\mZ_2 (\beta, t) = \frac{1}{Z^2} \sum_{m,n} e^{-\beta(E_m + E_n)} e^{i t(E_m-E_n)} ,
\label{sffdef2}
\eeq
where $Z (\beta+it)=\sum_n e^{-(\beta+it)E_n}$.
For a many-body chaotic system, $\mZ_2 (\beta, t)$ has distinct regions as a function of $t$. Starting from $t = 0$, it has a power law decay until a particular time $t = t_d$ (dip time) where the SFF attains its minima.  For the chaotic system with $N$ many body energy states, $t_d \sim \sqrt{N}$ \cite{Br_zin_1997}. After this point the connected part of the SFF starts to become important and the SFF starts to take the universal random matrix ensemble form manifested as a linear growth of SFF with time. This region is denoted as the `ramp'. The ramp ends at $t = t_{\texttt{plateau}} \sim N$, denoted as the plateau time or the Heisenberg time. This timescale scales as the inverse of the average level spacing of the system, thus after this point the discreteness of the energy spectrum becomes important. SFF saturates after $t_p$ and is denoted as the `plateau' \cite{Cipolloni2023, SierantPRL2020}. This linear ramp is denoted as a characteristic feature of a many-body chaotic interacting system \cite{cotler2017, liu2018} (see Fig.\ref{fig:sffGUE} where SFF is plotted for a random matrix drawn out of a Gaussian Unitary ensemble \cite{Brezin:1993qg, Meh2004, Cotler:2017jue}). On the contrary, the SFF in an integrable system does not show such ramp; the SFF in this case decays and then immediately plateaus.

In a non-interacting fermionic system, given a set of single particle eigenvalues $\epsilon_n$, the generalized partition function is given by
\beq
Z(\beta+ i t) = \prod_{n} \Big(1+ \exp(-(\beta+i t)\epsilon_n)\Big) ,
\eeq
where $n \in {0, \ldots, L-1}$, $L$ is the system size. Thus, for a symmetric eigenspectrum $(\epsilon_n \longleftrightarrow -\epsilon_n)$, the SFF (see \eqn{sffdef}) can be written as
\beq
\mZ_2 (\beta, t) = \prod_{\epsilon_n >0 } Z_2^{\epsilon_n}(\beta,t)= \prod_{\epsilon_n > 0} \frac{ \left( \cosh (\beta  \epsilon_n)+\cos (\epsilon_n t) \right)^2 }{\left( 1 + \cosh(\beta \epsilon_n)\right)^2} ,
\label{SFFZ2}
\eeq
which implies that the real-time behavior takes a highly convoluted form dependent on the frequencies of the single-particle energies.

From \eqref{sffdef2}, it may then appear that the time-averaged value is  $\sim Z(2\beta)/Z(\beta)^2$ on general grounds. Our results point out that even in an otherwise dense spectrum, the existence of topologically protected zero modes may render this asymptotic value insignificant and mask it with generalized oscillations. To elaborate this further, we now look into the features of SFF in some specific symmetry-protected topological models.

\section{Model} \label{model}

\begin{figure*}
    \centering
    \includegraphics[width=0.98\linewidth]{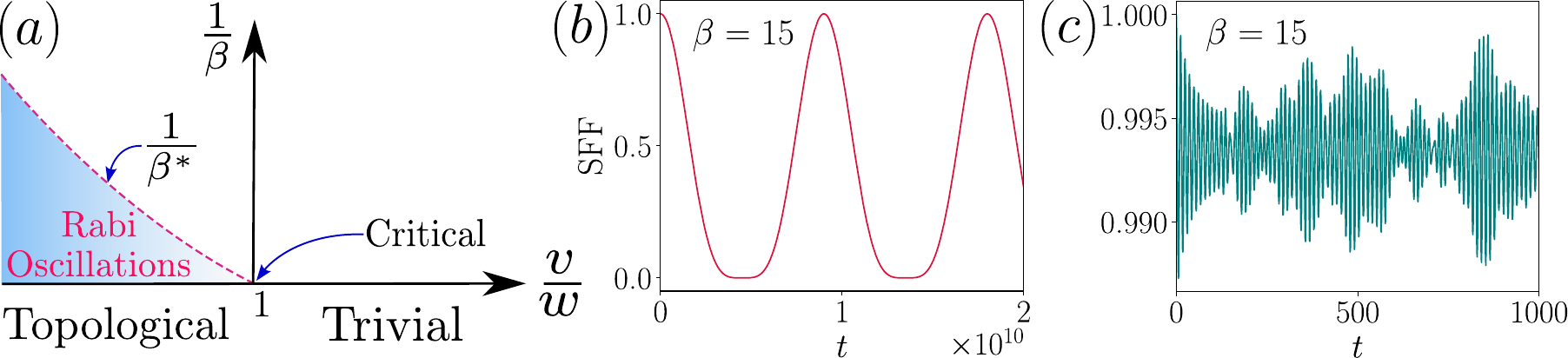}
    \caption{(a) A topological phase achieved in a Hamiltonian system (see \eqn{eqH} for SSH Model) is tuned by a parameter $v/w$. This sets an energy scale $T^* = \frac{1}{\beta^*}$ (shown by a dashed line) below which the system shows prominent oscillations (see (b), with $L=60, v/w = 0.5, \beta = 15$) in its SFF. (c) The same system in the trivial region ($ v/w = 2$) shows no signs of such oscillations. }
    \label{NTISSH}
\end{figure*}

To study the behavior of SFF in symmetry protected topological (SPT) phase, we first consider the paradigmatic SSH model~\cite{ssh} where spinless fermions hop in a one-dimensional chain via the following Hamiltonian: 
\beq
H_{\text{SSH}} = -\sum_{i} \big(vc^\dagger_{iA} c^\nd_{iB} + w c^\dagger_{iB} c^\nd_{i+1, A} + \text{h.c.}\big),\label{eqH}
\eeq
where $c^\dagger_{i\alpha}$ ($c^\nd_{i \alpha}$) is the fermionic creation (annihilation) operator at site $i$ for the orbitals $\alpha \equiv A, B$ and $v$, $w$ are the intra and inter-unit cell hopping strength respectively. The system has time-reversal and sub-lattice symmetry restricting it to the BDI symmetry class, which realizes general off-diagonal real matrices of the free fermion ten-fold classification \cite{Altland_PRB_1997, Agarwala_AOP_2017}, such that the system spectra is always symmetric about energy $E=0$. The bulk spectrum of the model is given by, $E(k) = \pm \sqrt{v^2 + w^2 + 2vw\cos{k}}$, where $k$ is the discrete momenta of the periodic chain. At half-filling (i.e., pinning the Fermi energy at $E_F = 0$), by tuning $v/w$, the system goes from one insulator to another insulator via quantum critical point $|v/w|=1$. The insulating phase in $|v/w|<1$ regime is topological and characterized by a non-trivial winding number of the bulk band, the corresponding open-chain hosts two close to zero energy modes at the boundary with an edge localization length $\xi= [\log(|w/v|)]^{-1}$. To see the feature in the SFF with an underlying symmetry-protected topology, we will always consider an open SSH chain so that the spectrum contains boundary-localized zero energy modes.

 In the trivial region ($|v| > |w|$), for all values of $\beta$, the SFF $\mZ_2 (\beta, t)$ is a superposition of all the single-particle energies (see \eqn{SFFZ2}), thus resulting in a convoluted noisy oscillation. This is rather uninteresting, as can be seen by the behavior of SFF in the trivial region with $v/w=2$ (\Fig{NTISSH}(c)). However, in the topological regime, an interesting behavior emerges when boundary modes dominate, whose energy $\epsilon_1 \rightarrow 0$ exponentially in system size. This results in a time scale $\sim \pi/\epsilon_1$ where the SFF first goes to zero and then oscillates with the same frequency, drowning all noisy behavior due to higher energy modes which are exponentially killed due to a finite $\beta$ (see \Fig{NTISSH}(b), $v/w=0.5$ topological region). These long-time oscillations can be understood as generalized Rabi oscillations, which we now explore further.

\subsection{Rabi Oscillations}

In general, given any initial state of a two-level system, the way the probability density of the state oscillates under time evolution is known as Rabi oscillations.
While generally discussed in context of a time-dependent perturbation, even for a constant perturbation switched on at $t=0$
the overlap of the unperturbed eigenstates, with the time evolved state oscillates with a characteristic frequency decided by the energy gap of the two level system\cite{Sakurai1993Modern, griffiths2018introduction}. We next show that the oscillations in the SFF can be understood as exactly  these oscillations for a generalized initial state.



We now discuss that the long-time oscillations of the SFF are, in fact, generalized Rabi oscillations, where the system locks between the boundary modes. Consider a wavefunction which is an equal superposition of {\it all} the many body basis states 
\beq
|\Psi\rangle = \frac{1}{2^L}\sum_{\{N_n\}} |\{N_n\} \rangle ,
\label{inistate}
\eeq
where $|\{N_n\} \rangle$ specifies the Fock state representation labelled by occupancies ($N_n = 0, 1$) of the single particle state $|\psi_n\rangle$ with eigen energy $\epsilon_n$. When  `quenched' with the Hamiltonian, its fidelity at a later time is
\beq
{\cal F}(t) = \langle \Psi | \Psi (t) \rangle \propto Z(it) .
\eeq
Thus, instances where $Z(t) \rightarrow 0$ are Rabi oscillations of a pure state under time evolution. At $\beta=0$, the fidelity behavior will be uncharacteristic because all $\epsilon_n$s will show convoluted oscillations. 
\begin{figure}
    \centering
     \includegraphics[width=0.99\linewidth]{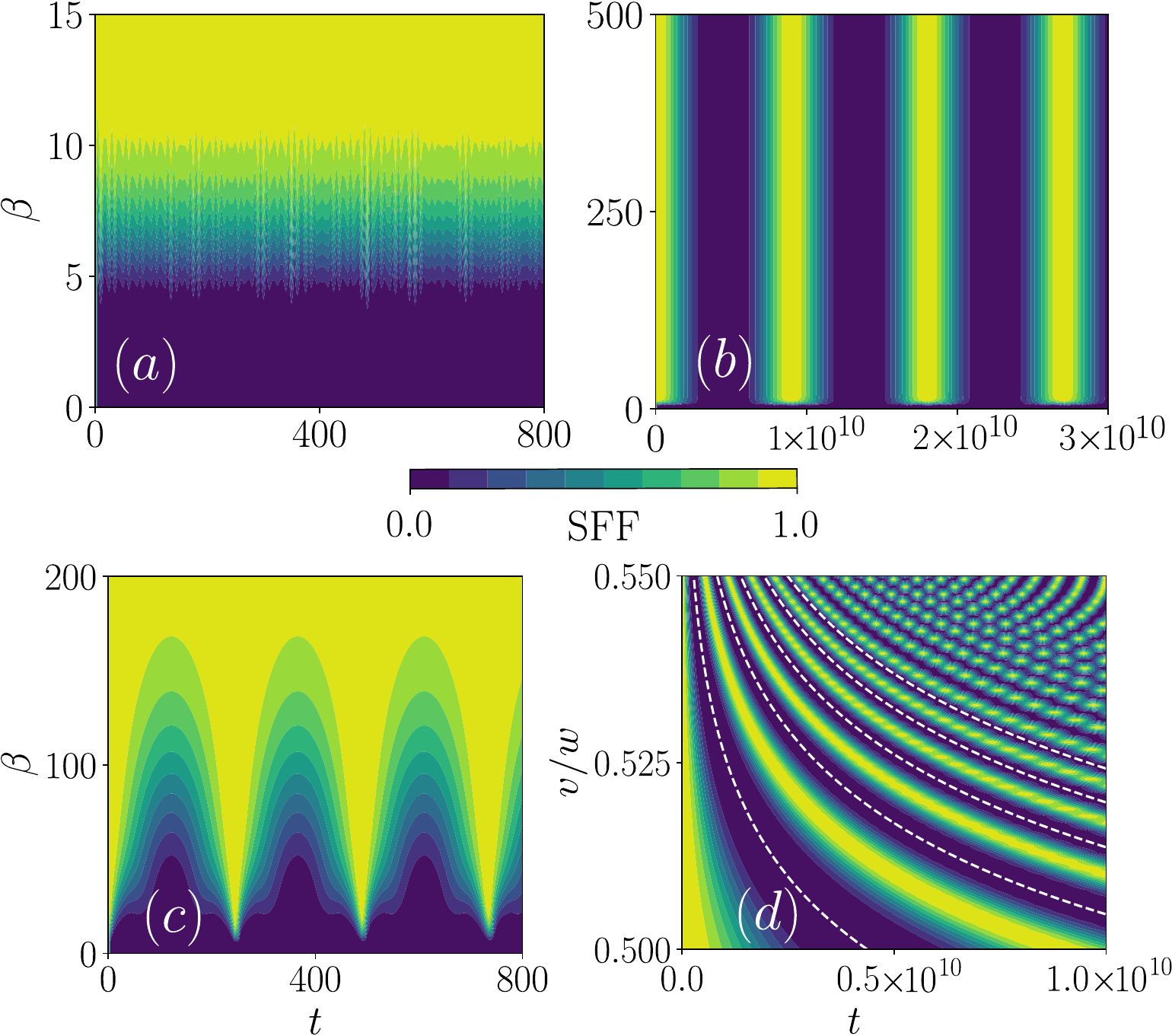}
     \caption{Contour plot of SFF in the $\beta-t$ plane for an open boundary SSH model (L = 60) in trivial regime (a) with $v=1, w=0.5$, in topological regime (b) with $v=0.5, w=1$ and at the critical point (c) with $v=0.5, w=0.5$.  (d) Behavior of SFF in the $v/w - t$ plane for $\beta = 30$. The dashed lines are analytical curves denoting the zeros of the SFF (see text).}
    \label{contourP}
\end{figure}However, at a finite temperature, the initial state (see \eqn{inistate}) can be generalized 
to 
\beq
|\Psi(\beta) \rangle \propto  \sum_{\{N_n\}} \exp(-\beta E_{\{N_n\}}/2) |\{N_n\} \rangle .
\eeq
When evolved in time, the corresponding fidelity is $Z(\beta + it)/Z(\beta)$, and thus SFF is just $|{\cal F}|^2$. 
Hence, SFF $\rightarrow 0$ are essentially the zeros of the Rabi oscillations of $|\Psi(\beta)\rangle$. Note that there is no external driving in the system; rather, the Hamiltonian evolution itself acts like a `drive' engineering the oscillations from the parent state $|\Psi(\beta) \rangle$. Given the states are normalized by the thermal occupancy factors, the states with the largest weights are where $N_n=1~\forall~ \epsilon_n<0$. In a periodic system, this isolates a single state; however, in an open one-dimensional SSH chain, when in topological phase -- this results in {\it four} states which are exponentially close in their many body energies. These correspond to different ways of occupying the (right and left) boundary modes ($\equiv |R\rangle, |L\rangle $). Thus, 
\bea
|\Psi(\beta, t) \rangle &\sim& \frac{1}{2} \Big( |\circ \circ\rangle + e^{-(\frac{\beta}{2}+it)\epsilon_a} |* \circ\rangle \\ &+& e^{-(\frac{\beta}{2}+it)\epsilon_b} |\circ *\rangle  +  |**\rangle   \Big) \notag \otimes | N_n=1~\forall~\epsilon_n<0\rangle
\label{effstate}
\eea
where $\epsilon_a, \epsilon_b$ ($\epsilon_b=-\epsilon_a$) represent the anti-bonding and bonding orbital combined out of the left and right edge states $(\{|b\rangle,|a\rangle\} = \frac{1}{\sqrt{2}} (|R\rangle \pm |L\rangle)$ and $*(\circ)$ represents their occupancies (vacancies) in  $|n_a,n_b\rangle$ basis. Because of the bulk gap $\Delta_g \sim |w-v|$ between the valence and conduction bands, other states get exponentially damped by $\sim \exp(-\beta \Delta_{g})$. This introduces a temperature scale $T^* \sim \frac{1}{\beta^*}=\Delta_g$, below which the Rabi oscillations are strong. For $T>\Delta_g$, these Rabi oscillations dissolve with the bulk signatures.  As is clear, the Rabi oscillation period is determined by edge mode energies $\sim \frac{\pi}{\epsilon_a}$. Unlike a single qubit, in this case, the SFF behaves as 
\beq \label{sshsffRabi}
\text{SFF}(t) \sim (1 +  \cos(\epsilon_at))^2.
\eeq 
Thus, the rise from the minima is $\propto t^4$ (see \Fig{HOTI}(d).

In \Fig{contourP}, we show the behavior of the SFF in the trivial, topological and critical point on the $\{\beta, t\}$ plane. We see that the zeroes of SFF disappear for small values of $\beta$ in the trivial phase, while in the topological phase, the zeroes persist even for high $\beta$. This denotes that the topological phase (see \Fig{contourP}(b)) has dominant oscillations.

From \eqn{sshsffRabi}, the oscillations have a time period $\sim \pi / \epsilon_a$. As for an SSH chain of length $L$ the zero energies scale as $\epsilon_a \sim \exp \left( -L \log \left| \tfrac{w}{v} \right| \right)$, the time period scales as $\sim \pi \exp \left(-L \log \left| \tfrac{v}{w} \right| \right) $. The SFF starts from its maxima at $t = 0$ and attains its first zero minima at half time period $\sim \pi \exp \left(-L \log \left| \tfrac{v}{w} \right| \right) $ (see \Fig{NTISSH}(b)). Thus, the subsequent minimas appear at the odd multiples of this half period; we denote these points as $t_{\texttt{Rabi}} \sim (2 k-1) \pi \exp \left(-L \log  \left| \tfrac{v}{w} \right| \right)$ with $k \in \mathbb{Z}^{+}$. Numerically the exact expression can be obtained as:
\begin{align}
	t_{\texttt{Rabi}} = 2.653 (2 k -1)\pi  \exp\left[-0.483 \, L \log \left( \left| \frac{v}{w} \right| \right)\right] .
\end{align}
This expression is consistent with the behaviour of the SFF in topological phase, as can be seen in \Fig{contourP}(d) when plotted on the ${v/w , t}$ plane.

\subsection{HOTI phase}

\begin{figure}[h!]
    \centering
    \includegraphics[width=0.99\linewidth]{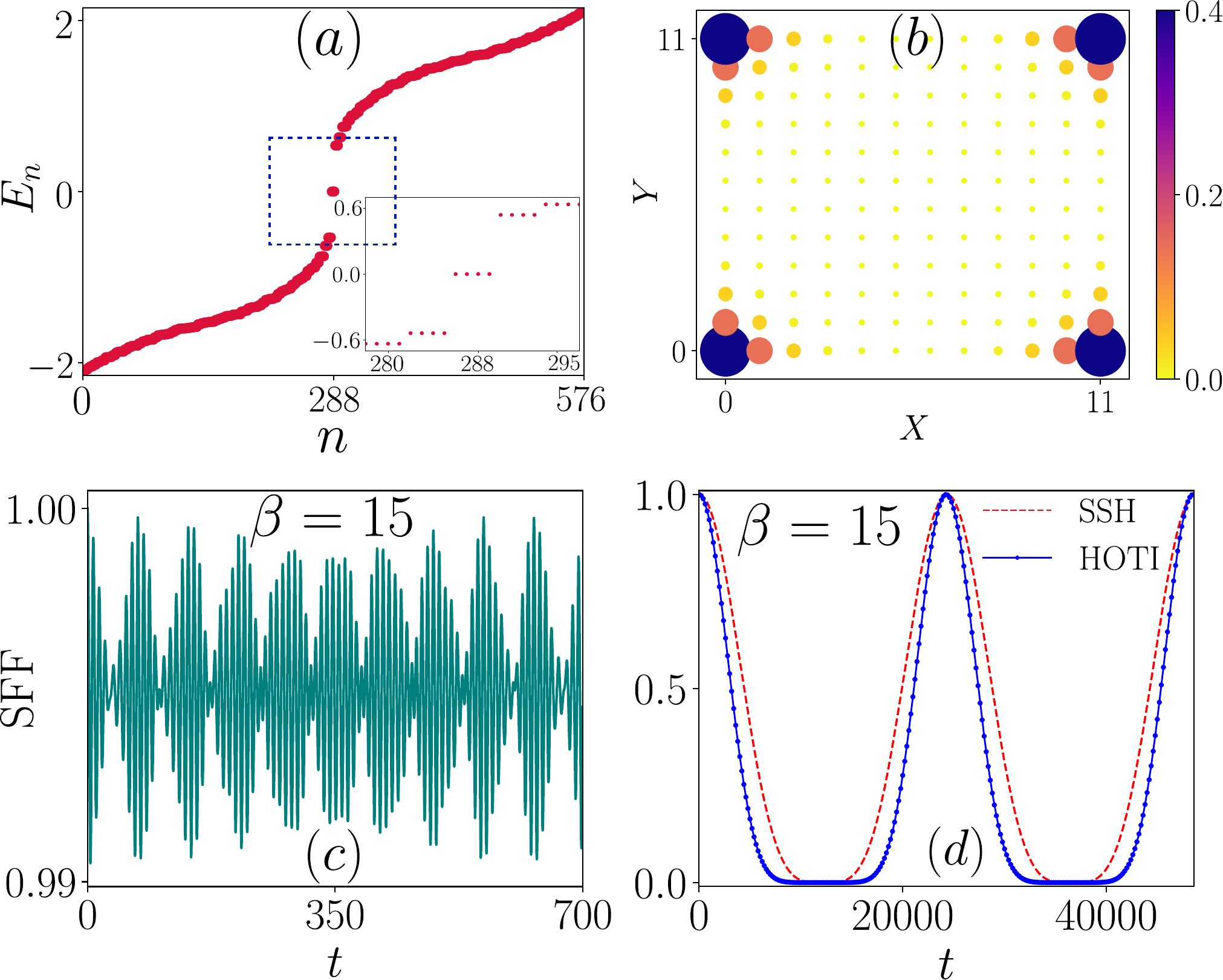}
    \caption{(a) One-particle energy spectrum of 2D HOTI model (see text) ($\gamma = 0.5$, $\lambda = 1$) with open boundary condition on a $12\times 12$ lattice. In the inset, the four topological zero energy modes are shown. (b) Local probability density of the zero modes showing that they are corner localized (c) Plot of SFF in the trivial regime ($ \gamma =1, \lambda=0.5$) (d) Generalized Rabi oscillations in the topological regime of HOTI (blue) ($\gamma = 0.5, \lambda = 1$) and SSH system (red) ($v=0.5154, w=1, L=24$). The rise of the SFF in the former is $\sim t^8$ compared to $\propto t^4$ in the latter. In both (c) and (d) SFF is calculated for $\beta = 15$.}
    \label{HOTI}
\end{figure}

In order to further investigate the behavior of SFF in the presence of topologically protected zero-dimensional boundary modes, we now focus on a HOTI model where spinless electrons on a square lattice host four corner states  \cite{benalcazar2017quantized}. The Bloch Hamiltonian of the model is of a four-band insulator given by,
\bea
H(k) = [\gamma &+& \lambda \cos(k_x)] \Gamma_4 + \lambda \sin(k_x) \Gamma_3 \nonumber \\ &+& [\gamma  +\lambda \cos(k_y)] \Gamma_2 + \lambda \sin(k_y) \Gamma_1,
\label{hotimodel}
\eea
where $\Gamma_i=-\tau_2\sigma_i$ for $i=1, 2, 3$, and $\Gamma_4= \tau_1\sigma_0$; $\sigma, \tau$ are Pauli matrices which act on the four orbitals of a unit cell. The model preserves both time-reversal and charge-conjugation symmetry, so it belongs to the BDI symmetry class, same as the SSH model. The bulk energies of the Hamiltonian is given by $E(k) = \pm \sqrt{2\lambda^2 + 2\gamma^2 + 2\gamma \lambda [\cos(k_x) + \cos(k_y)]}$, each of which is doubly degenerate. The gap in the energy bands closes at $|\gamma / \lambda| = 1$. At half-filling, when $|\gamma/\lambda|<1$, the overall charge density is essentially localized at the corners of the open square lattice  (see \Fig{HOTI}(a), (b)), resulting in a non-trivial bulk quadrupole moment of the insulator \cite{Kang_prb_2019}. On the other hand, when $|\gamma/\lambda|>1$, both corner states and quadrupole moment vanishes, and the insulator becomes trivial.

Evaluating the SFF, we again find noisy oscillations in the trivial regime and long-time oscillations in the topological regime (see \Fig{HOTI}(c), (d)). Here, the effective many-body state spans over $2^4$ states, which can be counted as occupancies of four boundary modes. The oscillations follow,
\beq
\text{SFF}(t) \sim (3 + 4 \cos(\epsilon_at) + \cos(2\epsilon_at))^2 ,
\eeq
where $\epsilon_a$ is the exponentially small energy scale close to zero. Interestingly, the rise from zero is now $\sim t^8$, reflecting the higher number of zero modes in the system (see \Fig{HOTI}). In fact, for a multipole topological insulator\cite{benalcazar2017quantized} with a general number of $2p$ zero modes, SFF would the scale $\propto t^{4p}$, in the topological phase for $T<\Delta_g$. This is one of the key results of our work. 

It is now natural to pose what is the fate of such Rabi oscillations in the presence of disorder, the question we answer next.

\section{Bulk Randomized SSH System}\label{bulkssh}

Motivated by random matrix theory (RMT), where the symmetry properties of random dense Hamiltonian matrices determine their chaotic signatures \cite{Dyson:1962es, prosen, chalker, Tezuka_PRB_2023, Kos_PRX_2018}, we introduce disorder to study the chaotic signatures in the SSH model as discussed in the previous section. 
While generic short-ranged disorder in topological phases given rise to a host of interesting phases and phase transitions \cite{Bala_prb_2017, Liu_pla_2018, Eric_science_2018, ste_opt_2020, Shi_prr_2021, Zuo_pra_2022}, here in order to simulate RMT chaos we introduce all to all hopping disorder simulating a `zero' dimensional chaotic system in conjunction with an underlying topological phase.

\begin{figure}
    \centering
    \includegraphics[width=0.95\linewidth]{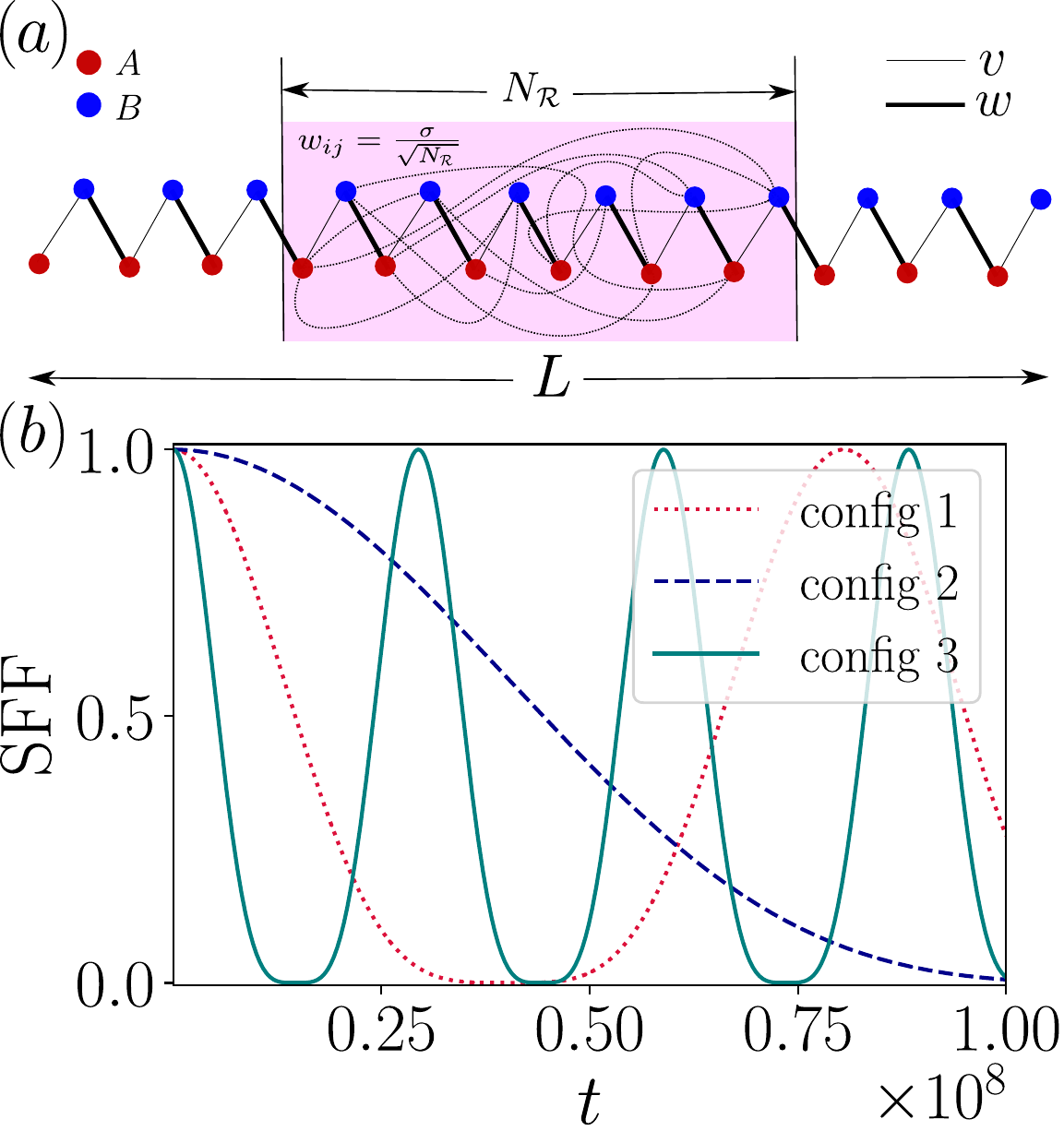}
    \caption{(a) Schematic diagram of a bulk-random SSH chain (see \eqn{eqrandSSH}), where symmetry-preserved all to all hopping disorder is introduced in the central bulk region $\cal{R}$ with $N_{\cal R}$ number of sites of a clean open SSH chain (see \eqn{eqH}) of size $L$. The disordered hopping strengths $w_{ij}$'s ($\{i,j\} \in {\cal R}$) are chosen from Gaussian distribution with scale parameter $\sigma /\sqrt{N_{\cal R}}$. (b) SFF in bulk-random SSH model ($L =60, \beta= 50, \sigma=0.1, N_{\cal R} = 30$) for different disorder configurations in topological regime ($v=0.5, w=1$) shows long-time oscillations with different time periods.}
    \label{brssh}
\end{figure}

To this effect, we mark a finite central region (excluding boundaries) in the bulk of the SSH chain (see \eqn{eqH}) as  $\equiv {\cal R}$, where such hopping disorder is introduced (see \Fig{brssh}). The microscopic Hamiltonian of such a system is,
\beq
H_{\text{R}} = H_{\text{SSH}}-\sum_{\{i,j\} \in \cal{R}} \big(w_{ij}\, c^\dagger_{iA} c^\nd_{jB} + \text{h.c.}\big).\label{eqrandSSH}
\eeq Here $w_{ij}$'s are chosen from a Gaussian orthogonal ensemble with a scale parameter $\frac{\sigma}{\sqrt{N_{\cal R}}}$ where $N_{\cal R}$ is the number of sites in the region ${\cal R}$ and $\{i,j\} \in {\cal R}$ (see \Fig{brssh}(a)). Since the disorder respects the sublattice character, given ${\cal R}$ excludes the edges and $\sigma \ll |w-v| $, every disorder configuration will retain topologically protected zero modes. The numerical support for the stability of the topological phase with disorder strength and the length of the randomized region is given in Appendix \ref{apndxA}, where we show for small enough disorder strengths, the system described in \eqn{eqrandSSH} has quantized polarization. As every random realization of the disordered SSH chain is individually topological, the individual SFFs show long-time oscillations similar to a clean topological SSH chain (see \Fig{brssh}(b)). Since in this work we have focused on disordered but quadratic systems where zero energy manifold plays a crucial role, we do not investigate the effects of unfolding and filtering, which is designed to decipher many-body bulk chaotic signatures \cite{unfolding2020PRE}.

\begin{figure*}
	\centering\includegraphics[width=0.95\linewidth]{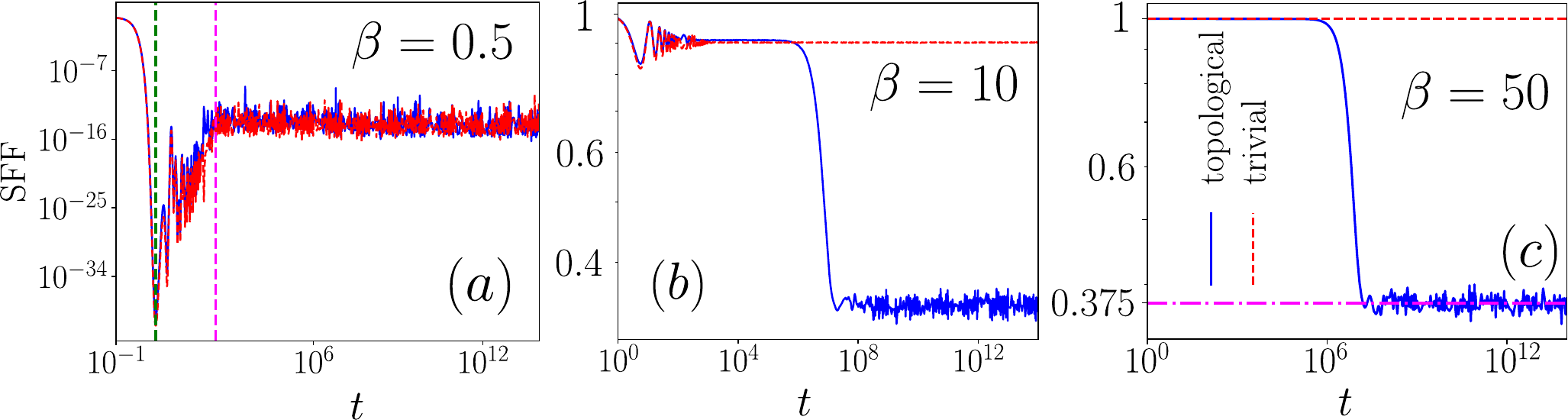}
	\caption{(a) The SFF (disordered averaged over 2000 configurations) in random SSH model ($L =60, \beta= 0.50, \sigma=0.01, N_{\cal R} = 30$) shows early time ramp in both topological ($v=0.5, w=1$) and trivial regimes ($v=1, w=0.5$). The green and magenta dotted vertical lines denote the dip-time and plateau time calculated from the toy model. (b) With increasing $\beta$ ($=10$), the initial dip starts to disappear and a new plateau forms at a late time. (c) At very high $\beta$ ($= 50$), the early-time ramp disappears; the late-time plateau in the topological SFF saturates around $3/8$. Here, $\Delta_g^{-1} \sim 2$ and $\epsilon_0^{-1} \sim 10^{7}$.) Note that trivial phase has saturated to a plateau value of unity.}
	\label{randomSSH}
\end{figure*}

\subsection{SFF Behavior}

We find that the disorder averaged SFF, irrespective of the topological character of the phase, has the following behavior as a function of time: (i) a dip, (ii) an exponential oscillation and then a linear ramp, and (iii) a late time saturation. The three regions are shown in \Fig{randomSSH}(a). As one decreases the temperature (i.e., increasing $\beta$), the exponential ramp starts to disappear (\Fig{randomSSH}(b)). For $\epsilon_0^{-1} \gg \beta > \Delta_g^{-1}$, the exponential ramp is almost entirely suppressed and the late time plateau value saturates around a new plateau value of $\sim 3/8 = 0.375$ instead of the usual plateau value $\sim \text{system size}^{-1}$ (see \Fig{randomSSH}). 

At short times, the {\it exponentially} oscillating behavior is characteristic of signatures of single particle chaos \cite{Winer:2020mdc, Liao:2020lac,Chen:2022hbi}
which is in contrast to many body chaos that has a distinct linear ramp right after the dip \cite{cotler2017, liu2018}. To delve into an understanding of the different features of SFF of the bulk-disordered SSH model, which is analytically intractable, one needs a simpler setting. To this end, we construct a minimal toy model which we discuss next.

\begin{figure}
	\centering
	\includegraphics[width=0.98\linewidth]{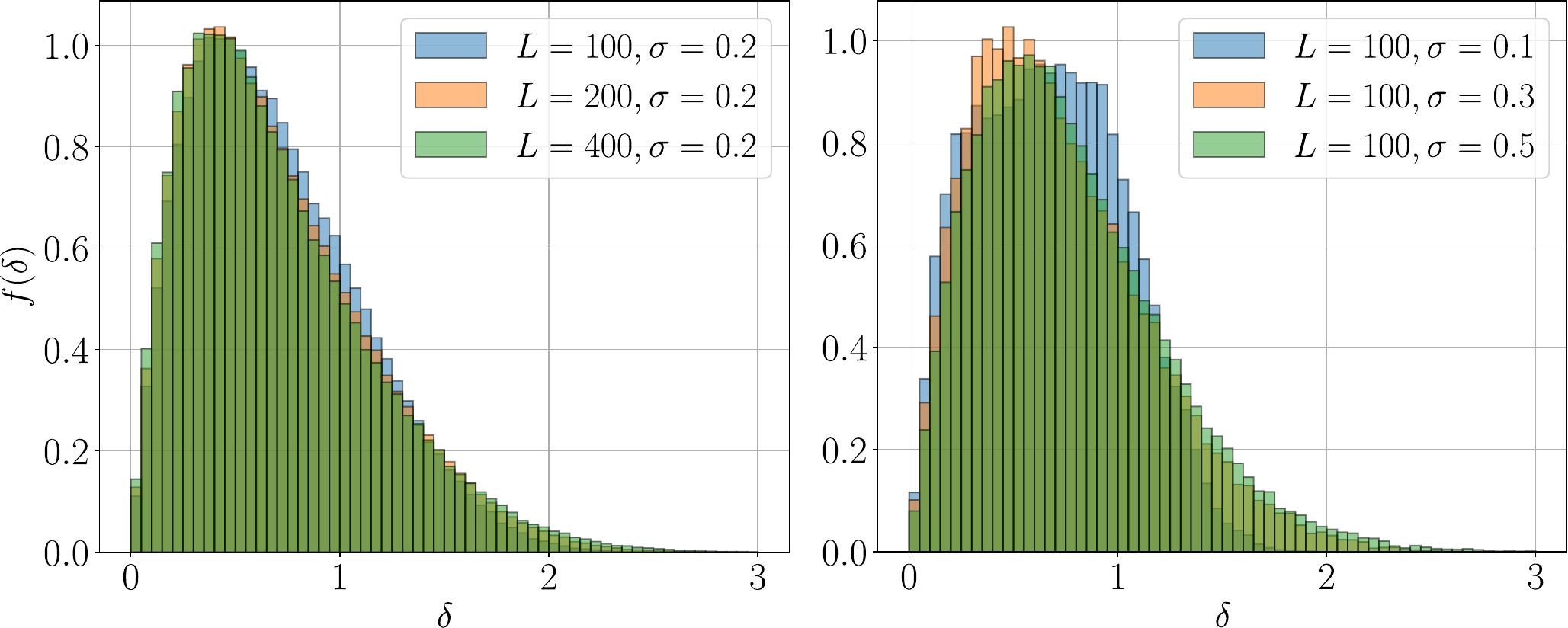}
	\caption{Random SSH single particle level spacing distributions: (a) for the same randomness but in different system sizes, (b) for the same system size but different randomness. Here $v=1, w=0.5$.}
	\label{SSHls}
\end{figure}

\subsection{Understanding using a toy system}

In order to capture the physics of the disordered SSH model, we first motivate the construction of an effective toy model. In \Fig{SSHls}, we plot the single particle nearest neighbor level spacings in the disordered SSH models, which show that they have a distribution similar to that of Wigner-Dyson (WD). This motivates us to construct a toy model using the eigenvalues from GUE of random matrices, which has WD distributed nearest neighbor level spacings. This is also pertinent given that the nature of the ramp arises due to the correlation between the level spacings of the spectrum~\cite{cotler2017}.

\begin{figure}[h!]
\centering
\includegraphics[width=0.95\linewidth]{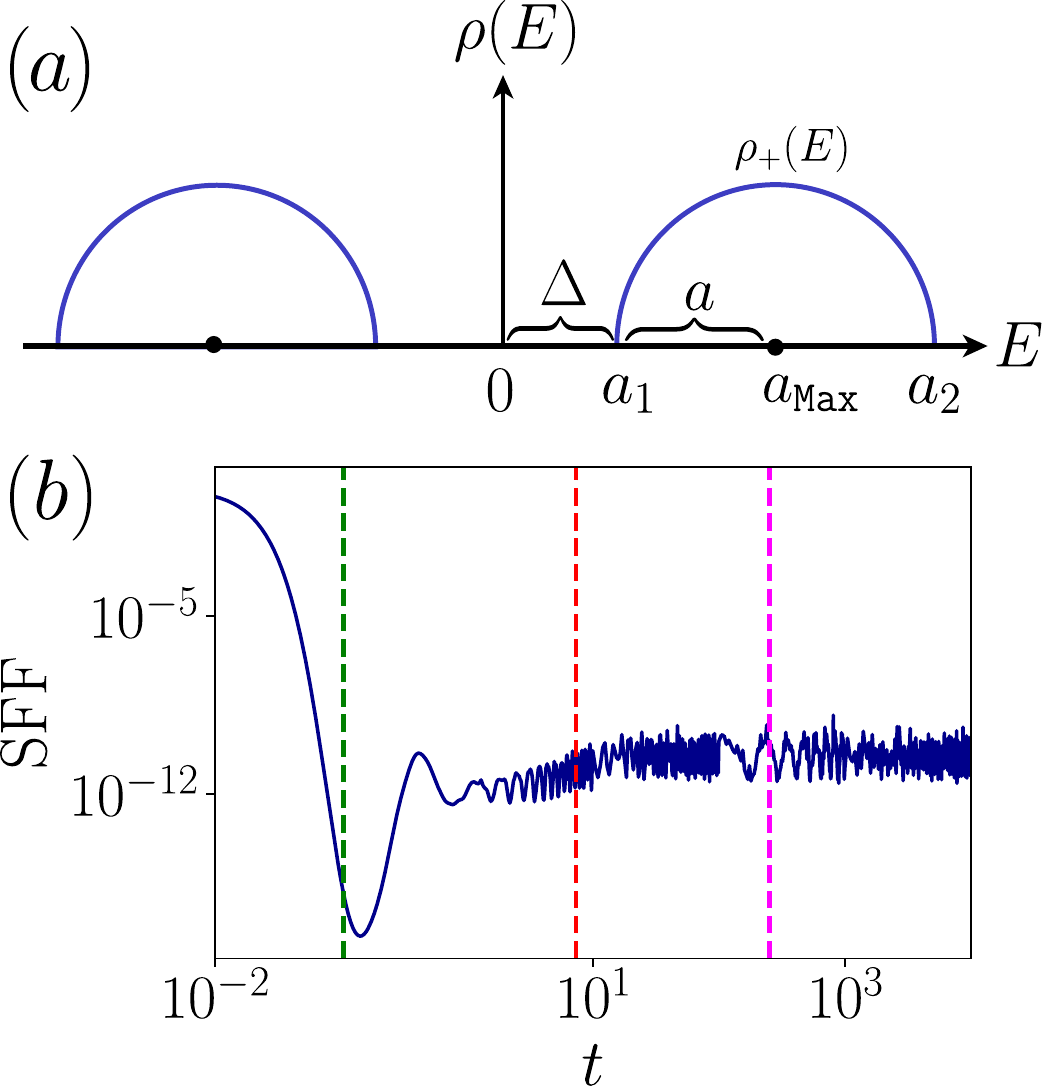}
	\caption{(a) Toy model density of states. (b) Toy model ensemble averaged SFF for $a_{\texttt{max}} = 30, \, a = 20, \, N = 20$. The green, red and magenta lines denote $t_{\texttt{dip}}, ~ t_{\texttt{crossover}}, ~ t_{\texttt{plateau}}$ respectively.}
	\label{toyDOS}
\end{figure}

The toy model consists of $N$ one-particle energy states taken from a GUE of random matrices which has semicircular $\rho(E)$ of radius $a$, centered around $a_{\texttt{max}} = a + \Delta$ (with $\Delta > 0$) and a corresponding negative $E$ copy (see \Fig{toyDOS}$(a)$). The positive energy branch of density of states (DOS) $\rho_{+} (E)$ is distributed between $a_1 = a_{\texttt{max}} - a$ to $a_2 = a_{\texttt{max}} + a$ with a spectral gap of $\Delta$. The ensemble-averaged $\rho_{+} (E)$ is given by
\begin{align}  
	\vev{\rho_{+} (E) } = \frac{2}{a^2 \pi} \sqrt{(E - a_1)(a_2 - E)} .
\end{align}

The ensemble-averaged SFF is then,
\beq
\vev{\mZ_2^R(\beta,t)} = \bigg\langle \exp \left(N \int dE \, \rho_+(E) \log Z_2^E(\beta, t) \right) \bigg\rangle  .
\eeq 
The angular brackets indicate the RMT averaging, where all moments of $\rho_+(E)$ contribute. At early times, the disconnected piece dominates, leading to $\sim \exp\big[\left( N \int dE \, \langle \rho_+(E) \rangle  \log Z_2^E(\beta,t) \right)\big]$. Under a high-temperature expansion, the system exhibits the exponential ramp imbued with short-time oscillations. The $\beta \rightarrow 0$ limit is given by, 
\begin{align}
    \vev{\mZ_2^R(0,t)} \approx  \frac{\exp \left(  8N \sum_{k=1}^{\infty} \frac{(-1)^{k+1}}{k^2 a t}J_1(k a t) \cos (k a_{\texttt{max}} t)   \right)}{16^N} .
    \label{eq:Z2Rexp}
\end{align}
In a microscopic model, $a_{\texttt{max}}$ is the energy scale where the bulk DOS peaks. This points out that the location of the dip is set by high energy scale $\pi/a_{\texttt{max}} \sim t_{\texttt{dip}}$, which is both system size independent and is unaffected by its topological properties (see \Fig{randomSSH}). The initial exponential oscillations lead to a linear ramp at $ t_{\texttt{crossover}} \sim L^{2/3}$. This is a result of the leading connected piece, i.e., $\rho_c^{(2)}(E_1, E_2)$ becoming dominant, just as in the case of many-body SFF \cite{Cotler:2017jue} (see Appendix \ref{apndxB1}). The ramp further leads to a saturation, which can be remarkably different if the underlying system is trivial vs. topological. More interestingly, the saturation physics is temperature dependent, as we now discuss.

At high temperatures ($T>|w-v|$), irrespective of topological features, the SFF saturates to a value dominated by the Kubo gap of the bulk spectra. This is essentially where all connected spectral correlations are featureless. For the minimal model, this predicts the emergence of a plateau at $t_{\texttt{plateau}} \sim 4N \pi$ (see Appendix~\ref{apndxB})and $\sim L$ for a bulk disordered SSH system. However, now as $T$ is reduced below the bulk gap scale ($T<|w-v|$) (see \Fig{NTISSH}), depending on whether we are in the trivial or the topological phase {\it another} saturation plateau appears.

This difference at low temperatures is due to the presence of zero energy modes ($\sim \pm \epsilon_0$), which in the clean limit leads to large time Rabi oscillations (see \Fig{NTISSH}). The effective SFF can be captured in the form $\sim Z_2^{\epsilon_0}(\beta,t) \mZ_2^R(\beta,t)$ (see \eqn{SFFZ2}) where at low temperatures $Z_2^{\epsilon_0}$ dominates and thus any single configuration, even with bulk disorder, will lead to oscillations after a time scale $t^* \sim \frac{1}{\epsilon_0}$. However, the configuration-averaged SFF shows an intriguing behavior, as is shown in \Fig{randomSSH} (b) and (c). At low temperatures, the system reaches a {\it different} saturation value at later times. This central result,  as we next discuss, is {\it not} governed by large-$N$ RMT results but is rather governed by a small-$N$ RMT chaos.

\subsection{Insights from Random Matrix Theory}

For the toy model discussed above with $N$ single particle positive energy states, the late time saturation value goes as $\sim 2^{-N}$ at $\beta = 0$, as can be seen directly from \eqn{SFFZ2}. As $\beta$ increases, for large enough $\beta$, the $\cosh{\beta \epsilon_n}$ factor dominates in \eqn{SFFZ2}, thus the plateau value approaches 1. However, for the bulk-random SSH chain, the SFF in the topological phase saturates at $\sim 3/8 = 0.375$, while the trivial phase SFF plateaus around 1  (see \Fig{randomSSH} (b) and (c)). This indicates that in the trivial phase, i.e., in the absence of zero energy boundary modes, the late time saturation value is usually dictated by large-$N$ RMT chaos. On the contrary, the late time saturation in the topological phase is an artifact of the chaos in the zero-energy manifold, which we denote as small-$N$ RMT chaos.

\begin{figure}
	\centering
	\includegraphics[width=0.98\linewidth]{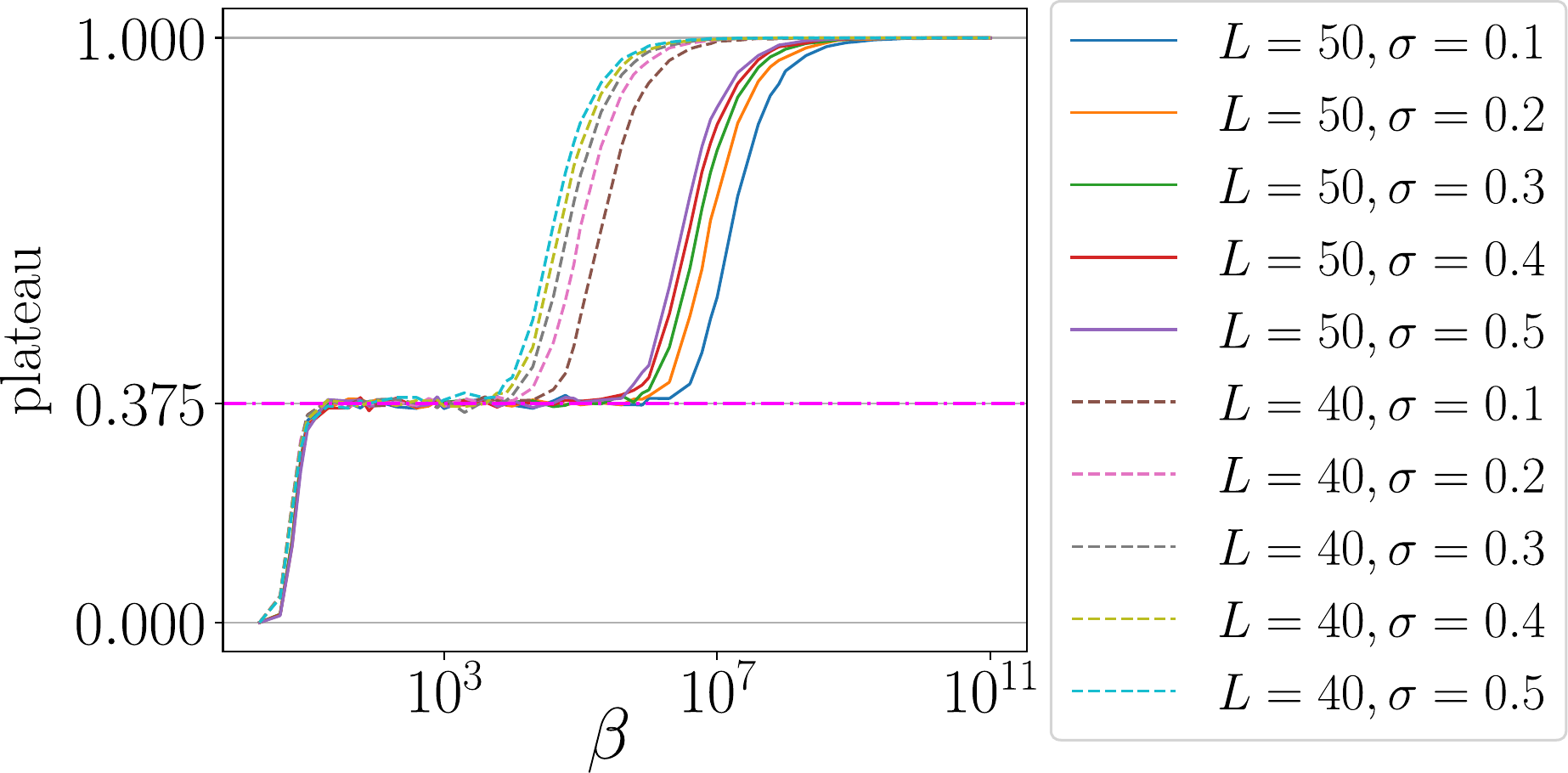}
	\caption{Late time ($t \sim 10^{12}$) saturation  values for different system sizes $L$ and $\sigma$ plotted with respect to $\beta$ (here $N_{\cal R} = L/2, v=0.5, w=1$). The SFFs are averaged over 2000 random configurations.}
	\label{plateau}
\end{figure}

Every peripheral disorder realization in the region ${\cal R}$ perturbs the edge modes, which are in turn protected by the $(w,v)$ scale of the underlying SSH Hamiltonian. Therefore, at low energies, i.e.~finite $\beta$, i.e., $ \beta > \Delta_g^{-1}$, the SFF gets dominated by RMTs reflecting the couplings within the zero energy manifold itself. For instance, in the clean SSH model, the effective Hamiltonian in the context of our example is a single qubit Hamiltonian $H_{\text{eff}} = \epsilon_0 \sigma_x$, written in the left and right edge state basis ($ |L\rangle, |R\rangle $). In presence of disorder, each realization effectively introduces a perturbation in this overlap between the basis states, i.e., the single qubit Hamiltonian now has an effective form: $H_{\text{eff}} = (\epsilon_0 + \lambda) \sigma_x$, where $\lambda$ can be considered to be drawn from a normalized Gaussian probability distribution $P(\lambda)$ distributed about zero mean with a variance $\sigma$. This perturbation thus results in a Gaussian distribution of energy eigenvalues $ \pm \epsilon_\lambda = \pm (\epsilon_0 + \lambda) $ centered around $\epsilon_0$ with variance $\sigma$. Thus, the  averaged $Z_2^{\epsilon_0}( \beta, t)$ becomes
\begin{align}
    \llangle Z_2^{\epsilon_0}( \beta, t) \rrangle &= \int_{-\infty}^\infty d\lambda\, P(\lambda)  \frac{ \left( \cosh (\beta  {\epsilon}_\lambda)+\cos ({\epsilon}_\lambda t) \right)^2 }{\left( 1 + \cosh(\beta {\epsilon}_\lambda)\right)^2} ,
    \label{eq:Z2Dis}
\end{align}
For the Gaussian distribution of $\epsilon_\lambda$ one obtains the long-time saturation value:
\beq
\lim_{t\rightarrow \infty} \llangle Z_2^{\epsilon_0}(\beta,t) \rrangle  \sim 3/8 ,
\eeq

when $ \epsilon_0 \ll T<|w-v|$ or equivalently $\Delta_g^{-1} \ll \beta \ll \epsilon_0^{-1}$.  Thus, this late-time behavior in the topological phase is fundamentally distinct from the trivial regime, where no such zero energy manifold exists, and the SFF just saturates to $1$. Note that, the long-time saturation value is independent of the system size and microscopic parameters, as indicated by the numerical results in \Fig{plateau}. Further lowering of temperature ($T < \epsilon_0$), even in the topological phase, drives the plateau towards unity as well. This is clear from \eqn{eq:Z2Dis}, as for large $\beta$ one obtains:
\beq
\llangle Z_2^{\epsilon_0}(\beta, t) \rrangle \simeq \int_{-\infty}^\infty d\lambda \, P(\lambda) = 1 .
\eeq
The analytical predictions exactly match our numerical results as shown in \Fig{randomSSH}(b). For further details of the SFF calculations, see Appendix \ref{apndxB2}.

This result can be further generalised for a topological phase containing $p$ pairs of symmetric zero modes: $\{ \epsilon_{0}^{(1)}, -\epsilon_{0}^{(1)} \}, \{ \epsilon_{0}^{(2)}, -\epsilon_{0}^{(2)} \}, \cdots , \{ \epsilon_{0}^{(p)}, -\epsilon_{0}^{(p)} \}$. In this case, under bulk disorder, each zero energy pair can be thought of arising from a single qubit effective Hamiltonian $H_{\text{eff}}^{(j)} = \epsilon_0^{(j)} \sigma_z + \lambda_j \sigma_x$ with energy eigenvalues $\epsilon^{(j)}_\lambda = \sqrt{ (\epsilon^{(j)}_0)^2 + \lambda_j^2}$. Here $\lambda_j$ is drawn from some normalised probability distribution $P_j(\lambda_j)$. Under the assumption that every zero-energy pair fluctuations are mutually independent, the effective Hamiltonian of the zero energy manifold is then:
$$ H_{\text{eff}}^0 = \bigotimes_{j=1}^{p} H_{\text{eff}}^{(j)} . $$
Then for a Gaussian distribution of $\epsilon^{(j)}_\lambda$ centered around $\epsilon_0^{(j)}$, the averaged zero-energy SFF is: 
\begin{align}
    \llangle \prod_{j=1}^{p} Z_2^{\epsilon^{(j)}_0}( \beta, t) \rrangle 
    \xrightarrow[]{t \to \infty} (3/8)^{p} .
\end{align}
This is the long-time plateau value of the SFF for the topological phase in the regime $\Delta_g^{-1} \ll \beta \ll \epsilon_0^{-1}$. Note that the bulk-disordered SSH model is the $p=1$ case, as it contains only one pair of zero energy modes. We note, however that effective nature of hybridization within the effective zero-energy manifold may  depends on the system and symmetries of the microscopic model.

\section{Conclusions} 
\label{concl}

In this work, we have shown that underlying topological order has important implications for chaotic signatures in quantum systems. Our analysis establishes that the emergence of zero energy boundary modes fundamentally changes the late-time behavior of SFF, a versatile tool that has been critical to diagnose chaos and thermalization in a host of systems. In \Sect{SFF}, we explored the idea of spectral form factor (SFF), its general features and how it is evaluated for a non-interacting system. We then discussed that in a clean topological system, SFF shows oscillations akin to generalized Rabi oscillations with features characteristic of the underlying topological properties. In particular, we study the behavior of SFF in the SSH and HOTI model, where the system hosts zero-dimensional edge modes (\Sect{model}). 
 In order to study the interplay with signatures of single particle chaos we introduce disorder in such a way that the edge modes remain intact. 
In particular, we introduce symmetry-preserved all-to-all hopping disorder in the bulk of an SSH chain (\Sect{bulkssh}).  Interestingly we find that bulk disorder alters the late-time SFF plateau in a topological phase. Gathering intuition from both from both numerical results and analytical calculations for effective toy models, our study uncovers the physics that the late time plateau is in fact determined by RMT behavior within the zero energy manifold. 

 While our work has investigated the role of non-interacting topology on the SFF, we have restricted to systems where the topological manifold provides zero modes such as in SSH or in HOTI systems. Here the separation of scales between the boundary manifold and bulk manifold is relatively clear. Even within symmetry protected topological phases, higher dimensional topological phases such as Chern insulators or topological insulators \cite{Hasan_RMP_2010, asboth2016short}, it may be interesting to explore how this physics changes. Here the eigenpectrum related to the boundary smoothly merges with the bulk spectrum thus making the physics further interesting.  Another natural question to pose is the role of both interactions and long-range topological order. Under both repulsive and attractive perturbative interactions, the SSH model is known to be stable and retains four-fold degenerate boundary modes in its many-body spectrum \cite{kohmotoPRB81,abhishodhPRL23,mondal2023symmetryenriched}. Thus, we expect the SFF to have long-time oscillations as long as the interaction does not close the bulk gap in the system. However, an elaborate study of SFF in interacting SSH and higher dimensional topological phases is an exciting future prospect.

\section{Acknowledgement}

A.S and S.P acknowledge funding from IIT Kanpur. A.A. acknowledges support from IIT Kanpur Initiation Grant IITK/PHY/2022010. D.D., \& A.S. acknowledge support by the Max Planck Partner Group grant MAXPLA/PHY/2018577. D.D. acknowledges support from MATRICS grant {{SERB/PHY/2020334}}. We acknowledge fruitful discussions at the ``Discussion Meeting on Non-equilibrium Correlated Systems" held in HRI, Allahabad.

\appendix

\section{Stability to Disorder}\label{apndxA}

\begin{figure}[h!]
    \centering\includegraphics[width=1\linewidth]{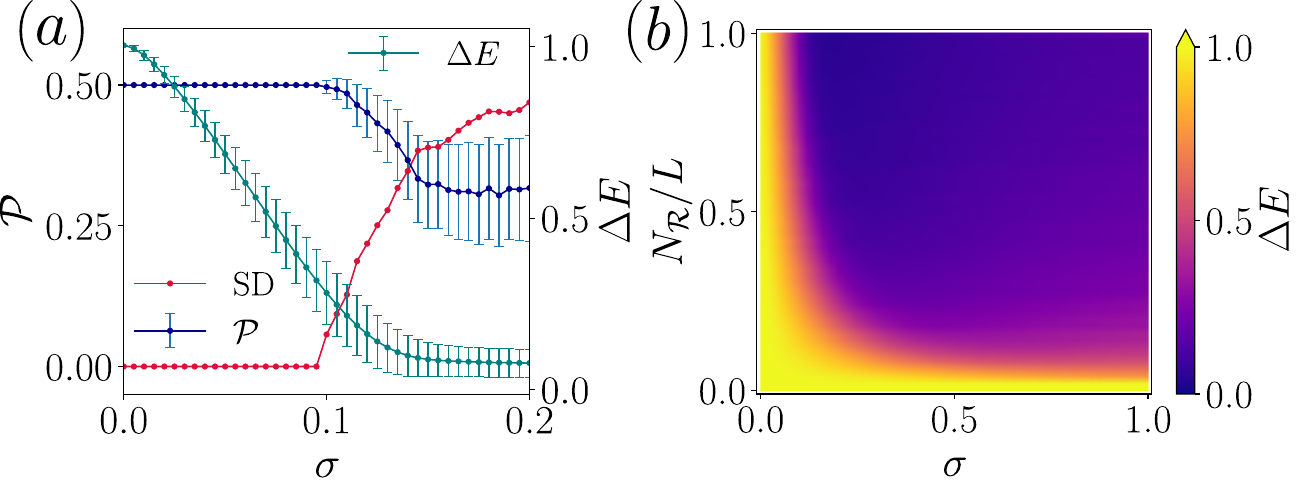}
    \caption{(a) The polarization $\mathcal{P}$ shows the stability of the topological phase up to a critical $\sigma$ for the SSH Hamiltonian with $90\%$ bulk region having all-to-all random hopping. The stability can also be seen from the standard deviation (SD) of the polarization, which is clearly related to the closing of the bulk gap ($\Delta E$) in the system. (b) Contour plot of the bulk gap with the disorder strength $\sigma$ and the fraction of random sites $N_{\mathcal{R}}/L$. For both the plots, the system size is $L=100$,  $v=0.5$, $w =1$ and all the data are averaged over $1000$ random configurations.}
    \label{fig: pol}
\end{figure}

In any topological phase, the presence of a bulk gap and underlying symmetries protects the boundary modes under perturbative disorder. In our work, directed towards studying the role of topological boundary modes in SFF, here we show that even in the presence of all-to-all random hopping (drawn from a Gaussian distribution with zero mean and variance $\sigma$) in a finite region $\cal R$, the boundary modes remain stable until a finite $\sigma$.  We study \textit{polarization}, which is the real space representation of the topological invariant (winding number) in $1$D systems~\cite{Resta_PRL_1998}. The polarization is defined as
\begin{equation}\label{eq:pol}
    \mathcal{P} = \frac{1}{2\pi} \text{Im} \bigg\{ \text{Tr}\Big[\ln\big(\hat{P}D\hat{P}\big)\Big]\bigg\}~~~ \text{mod $1$},
\end{equation}
where $\hat{P} = \sum_{E_n\leq E_F} \ket{\psi_n}\bra{\psi_n}$ is the ground state projection operator with $\ket{\psi_n}$ being the single-particle eigenstate corresponding to the energy eigenvalue $E_n$ and $E_F$ is the Fermi energy of the system. The position ($x_i$) of all the lattice sites (for system size $L$ there are $N=L/2$ unit-cells) are compactified and exponentiated to give the operator $D = \text{diag}\big[e^{2\pi x_{i\alpha}/N}\big]$, where $x_{i\alpha}= (i-1)$ for $i^{th}$ unit cell and both $\alpha=A,B$ sublattices. In the absence of disorder, in the topological regime of SSH Hamiltonian ($|v/w|<1$), the polarization $\mathcal{P}$ is quantized to $0.5$, while it is zero for the trivial insulating phase($|v/w|>1$). Furthermore, in \Fig{fig: pol}(a) we plot the polarization (see eqn.~\ref{eq:pol}) of the system in the topological regime ($v/w = 0.5$) as a function of bulk disorder $\sigma$ in $90\%$ $(N_{\cal R}/L=0.9)$ of bulk sites. The system shows quantized polarization as well as a finite bulk gap $\Delta E$ up to a finite, $\sigma$, confirming that the system's topological properties remain stable even when bulk disorder randomizes the bulk spectra. Furthermore, the stability increases as the number of bulk sites disordered is reduced; see \Fig{fig: pol}(b) where the bulk gap is plotted as a function of $N_{\cal R}/L$ and $\sigma$.

\section{Additional Details on SFF Calculations}\label{apndxB}

\subsection{Toy model for bulk disorder}\label{apndxB1}

Here, we go through a detailed analysis of the bulk-disordered toy model described earlier.

\begin{widetext}

	The many-body SFF when ensemble averaged is given by $\vev{Z^{R}_{2} (\beta, t)}$ which is
	\begin{equation} 
		\begin{split}
			&\vev{\mZ^R_2 (\beta, t)} = \vev{\exp \left( N \int \td E \, \rho_{+} (E) \, \log Z^{E}_2 \right)}
			= \sum^{\infty}_{n = 0} \frac{1}{n!} N^{n} \int \left(\prod_{i = 1}^{n} \td E_i \right) \vev{\rho_{+} (E_1) \ldots \rho_{+} (E_n)} \prod_{j = 1}^{n} \log Z^{E_j} (\beta, t), \label{intZ2R}
		\end{split}
	\end{equation}
	where a typical $n$-point density correlator has the form
	\begin{equation}
		\begin{split} \label{rho n point}
			&\vev{\rho_{+}(E_1) \ldots \rho_{+}(E_n)} \equiv \vev{\rho^{(n)}} = \prod_{i=1}^{n} \vev{\rho^{(1)}_i }  +  \sum_{\{n_i,m\}} A^{\{n_i,m\}} \prod_{i=1}^{k} \rho_c^{(n_i)} \prod_{j}^{m} \vev{\rho^{(1)}_j} 
			\text{ with: } \sum_{\substack{i=1 \\ n_i>1}}^{k} n_i + m = n; ~ \sum_{i}^{k} n_i \neq 0 .
		\end{split}
	\end{equation}
	Here, the first term denotes the disconnected piece, $A^{\{n_i,m\}}$ stands for the coefficient arising from the permutations of the indices, $\rho_c^{(n_i)}$ denotes the completely connected piece of the $n_i$-point correlation and $\rho^{(1)}_j \equiv \rho_{+} (E_j)$.

 Using the explicit form for the joint distributions, as given through determinant of the kernel \cite{Brezin:1993qg}, we find : 
	\begin{align}
		\prod_{i=1}^{n} \vev{\rho^{(1)}_i } &= {\cal O}( 1 ), ~~
		\prod_{i=1}^{k} \rho_c^{(n_i)} \prod_{j}^{m} \vev{\rho^{(1)}_j}  = {\cal O}\left( N^{m-n} \right) = {\cal O}\left( N^{-\sum_i n_i } \right) .
	\end{align}
	Since $\text{min}( n_i ) \geq 2$, the second term is always at least ${\cal O}(1/N^2)$ suppressed compared to the fully disconnected one. Of course, after evaluating the integrals in \eqref{intZ2R}, there are non-trivial time growths associated with various pieces. At early times, $t \ll N$, we can ignore these time dependencies and the dominant contribution, at large $N$, comes from the fully disconnected DOS correlator. Therefore, the sum over $n$ in \eqn{intZ2R} re-exponentiates, and  we find the early time, ensemble-averaged SFF to be:
	\begin{equation}
		\begin{split}
			&\vev{\mZ^R_2 (\beta, t)} 
			\simeq \exp \left( N \int \td E \, \vev{\rho_{+} (E)} \, \log Z^{E}_2 \right)  = \exp \left( N \int_{a_1}^{a_2} \td E \,  \frac{2}{a^2 \pi} \sqrt{(E - a_1)(a_2 - E)} \, \log Z^{E}_2 \right) .
		\end{split}
	\end{equation}
	At high temperatures, $\beta = 0$, the integral can be evaluated exactly in terms of Bessel function:
	\begin{equation}  \label{Z2R early t}
		\vev{\mZ^R_2 (0, t)}  \simeq \frac{1}{16^N}\exp \left(  8N \sum_{k=1}^{\infty} \frac{(-1)^{k+1}}{k^2 a t}  J_1(k a t)  \cos (k a_{\texttt{max}} t)  \right) .
	\end{equation}
	In the above expression, the higher frequency oscillations $ \cos (k a_{\texttt{max}} t)$ are suppressed as $k^{-2}$. Thus, the $k=1$ component dominates and produces the dip at the first oscillation minima: $t_{\texttt{dip}} = \pi /  a_{\texttt{max}} $.
	
	To see the dynamics at an intermediate time, we look at the connected component of, $\vev{\rho_{+} (E_1) \rho_{+} (E_2)}$ denoted as $\rho_c^{(2)} (E_1, E_2)$. In terms of sine kernel~\cite{cotler2017}, from \eqn{intZ2R} one obtains:
 
	\begin{equation}
		\begin{split}
			&\frac{1}{2!} N^{2} \int \td E_1  \td E_2 \rho_c^{(2)} (E_1, E_2) \prod_{j = 1}^{2} \log Z^{E_j}_2 = -\frac{ N^{2} }{2!} \int \td E_1  \td E_2  \frac{ \sin^2 \left( N \pi ( E_1 - E_2 ) \right) }{ N^2 \pi^2 ( E_1 - E_2 )^2 } \prod_{j = 1}^{2} \log Z^{E_j}_2  \\
			&= -\frac{2}{ \pi} \left(   N \pi \log^2 (4)   -  \log(4)  \sum_{k=1}^{\infty} \frac{(-1)^{k+1}}{k} \left(  4 N \pi - k t  \right)  \cos (k a_{\texttt{max}} t) \Theta \left(  \frac{4 N \pi}{k} - t  \right) + \right. \\
			&\left.   \sum_{k,l=1}^{\infty} \frac{(-1)^{k+l}}{k l}  \left[ \left(  4 N \pi - (k + l) t  \right)  \cos (|k-l| a_{\texttt{max}} t) \Theta \left(  \frac{4 N \pi}{(k + l)} - t  \right)  + \left(  4 N \pi - |k-l| t  \right)  \cos ( (k + l) a_{\texttt{max}} t) \Theta \left(  \frac{4 N \pi}{|k-l|} - t  \right)  \right]    \right) .
		\end{split}
	\end{equation}
	We see the 2-point connected piece generates linear terms of $\left(  4 N \pi - k t  \right)$ up to $t = 4 N \pi / k $. Thus, the $k=1$ component gives the longest surviving linear piece with the plateau time $t_{\texttt{plateau}} = 4 N \pi$ (see \Fig{toyDOS}$(b)$).
\end{widetext}
Comparing the disconnected and connected pieces for $n = 2$, one can estimate the time $t_{\texttt{crossover}}$ at which the connected piece starts to dominate over the disconnected piece. Using the asymptotic form of Bessel functions, one finds,
\begin{equation}
	N^{2} \frac{J_1(t_{\texttt{crossover}})}{t_{\texttt{crossover}}} \sim N ~~ \Rightarrow ~~ t_{\texttt{crossover}} \sim \mathcal{O} (N^{2/3}) .
\end{equation}
So one sees a linear ramp starting from $t_{\texttt{crossover}}$ up to $t_{\texttt{plateau}}$, and then the SFF plateaus.

As we see from the $n=2$ case, the surviving piece in long-time average is $\sim \mathcal{O}(N)$ for the connected piece, while the disconnected piece has a long-time averaged value of $\sim \mathcal{O} (N^2)$. Thus, in all $n$-point correlations, the disconnected piece gives the dominant long-time averaged contribution. From \eqn{Z2R early t}, we see in the long-time average the oscillations die down, thus, the plateau value becomes:
\begin{equation}
	\vev{\mZ^R_2 (0, t \to \infty)} \sim e^{-N} .
\end{equation}

\subsection{Late time SFF dynamics due to fluctuating zero modes:}\label{apndxB2}

In the 1D SSH model, there are two zero modes with energy $\pm \epsilon_0$ at the two ends of the lattice chain. When disorder is introduced in the bulk,  the zero modes also fluctuate. To model the spectrum of these two fluctuating zero modes, we consider a Gaussian energy distribution centered around $\epsilon_0$ with the variance $\sigma$. Thus, the ensemble-averaged zero energy contribution to the SFF is:
\begin{align}
	\llangle Z^{\epsilon_0}_2 (\beta, t) \rrangle &= \int \td \epsilon \frac{1}{\sigma\sqrt{2 \pi}} e^{-\frac{(\epsilon - \epsilon_0)^2}{2 \sigma^2}} \left(\frac{ \cosh (\beta \epsilon)+\cos (\epsilon t) }{ \cosh (\beta \epsilon)+1} \right)^2 .
\end{align}
For $\beta = 0$ (i.e, $\beta \ll \epsilon_0^{-1}$) we have,

\begin{equation}
	\begin{split}
		& \llangle Z^{\epsilon_0}_2 (0, t)\rrangle = \frac{1}{\sigma\sqrt{2 \pi}} \int \td \epsilon  e^{-\frac{(\epsilon - \epsilon_0)^2}{2 \sigma^2}} \left(\frac{ 1+\cos (\epsilon t) }{ 2 } \right)^2   \\
		&= \frac{3}{8} + \frac{1}{2} e^{-t^2 \sigma^2 /2}\cos (\epsilon_0 t) + \frac{1}{8} e^{-2 t^2 \sigma^2 }\cos (2 \epsilon_0 t) .
	\end{split}
\end{equation}

At large $t$, the oscillations will average to zero, thus leading to a plateau value of $3/8 = 0.375$.

To calculate the dip point, we can find the first minima of $\llangle Z^{\epsilon_0}_2 (0, t)\rrangle$ (ignoring the $\cos(2 \epsilon_0 t)$) term and arrive at the transcendental equation:
\begin{equation}
	\begin{split}
		\partial_t \llangle Z^{\epsilon_0}_2 (0, t)\rrangle = 0 ~ \Rightarrow ~ \tan (\epsilon_0 t) = -\frac{t \sigma^2}{\epsilon_0} .
	\end{split}
\end{equation}
When $\sigma \ll \epsilon_0$, in the above expression $\tan (\epsilon_0 t_{\texttt{dip}}) \to 0$, thus:
$t_{\texttt{dip}} \to \frac{\pi}{\epsilon_0}$. 
In this case, the exponential $\sim e^{-t^2 \sigma^2 /2}$ decay does not sufficiently suppress the $\cos (\epsilon_0 t)$ oscillations, thus one sees an oscillatory dip-ramp and plateau. 
For $\sigma \gg \epsilon_0$, $\tan (\epsilon_0 t_{\texttt{dip}}) \to -\infty$, hence:
$t_{\texttt{dip}}  \to \tfrac{\pi}{2 \epsilon_0}.$ In this case, the exponential decay dominates. Thus, one only sees a dip and plateau, without any ramp.
In the actual random SSH model, one has both the random excited states and fluctuating zero modes, i.e., $\vev{\mZ^{\texttt{Full}}_2 (\beta, t)} = \vev{\mZ^R_2 (\beta, t)} \llangle Z^{\epsilon_0}_2 (\beta, t) \rrangle$. Thus, when $\epsilon_0^{-1} \gg \beta > \Delta^{-1}$, the early-time dip-ramp due to $\vev{\mZ^R_2 (\beta, t)}$ starts to get suppressed, and one starts to see the late time SFF dynamics of $\llangle Z^{\epsilon_0}_2 (\beta, t) \rrangle$ (see Fig. \ref{randomSSH}).


\begin{thebibliography}{54}%
\makeatletter
\providecommand \@ifxundefined [1]{%
 \@ifx{#1\undefined}
}%
\providecommand \@ifnum [1]{%
 \ifnum #1\expandafter \@firstoftwo
 \else \expandafter \@secondoftwo
 \fi
}%
\providecommand \@ifx [1]{%
 \ifx #1\expandafter \@firstoftwo
 \else \expandafter \@secondoftwo
 \fi
}%
\providecommand \natexlab [1]{#1}%
\providecommand \enquote  [1]{``#1''}%
\providecommand \bibnamefont  [1]{#1}%
\providecommand \bibfnamefont [1]{#1}%
\providecommand \citenamefont [1]{#1}%
\providecommand \href@noop [0]{\@secondoftwo}%
\providecommand \href [0]{\begingroup \@sanitize@url \@href}%
\providecommand \@href[1]{\@@startlink{#1}\@@href}%
\providecommand \@@href[1]{\endgroup#1\@@endlink}%
\providecommand \@sanitize@url [0]{\catcode `\\12\catcode `\$12\catcode
  `\&12\catcode `\#12\catcode `\^12\catcode `\_12\catcode `\%12\relax}%
\providecommand \@@startlink[1]{}%
\providecommand \@@endlink[0]{}%
\providecommand \url  [0]{\begingroup\@sanitize@url \@url }%
\providecommand \@url [1]{\endgroup\@href {#1}{\urlprefix }}%
\providecommand \urlprefix  [0]{URL }%
\providecommand \Eprint [0]{\href }%
\providecommand \doibase [0]{http://dx.doi.org/}%
\providecommand \selectlanguage [0]{\@gobble}%
\providecommand \bibinfo  [0]{\@secondoftwo}%
\providecommand \bibfield  [0]{\@secondoftwo}%
\providecommand \translation [1]{[#1]}%
\providecommand \BibitemOpen [0]{}%
\providecommand \bibitemStop [0]{}%
\providecommand \bibitemNoStop [0]{.\EOS\space}%
\providecommand \EOS [0]{\spacefactor3000\relax}%
\providecommand \BibitemShut  [1]{\csname bibitem#1\endcsname}%
\let\auto@bib@innerbib\@empty
\bibitem [{\citenamefont {Srednicki}(1994)}]{Srednicki_PRE_1994}%
  \BibitemOpen
  \bibfield  {author} {\bibinfo {author} {\bibfnamefont {Mark}\ \bibnamefont
  {Srednicki}},\ }\bibfield  {title} {\enquote {\bibinfo {title} {Chaos and
  quantum thermalization},}\ }\href {\doibase 10.1103/PhysRevE.50.888}
  {\bibfield  {journal} {\bibinfo  {journal} {Phys. Rev. E}\ }\textbf {\bibinfo
  {volume} {50}},\ \bibinfo {pages} {888--901} (\bibinfo {year}
  {1994})}\BibitemShut {NoStop}%
\bibitem [{\citenamefont {D'Alessio}\ \emph {et~al.}(2016)\citenamefont
  {D'Alessio}, \citenamefont {Kafri}, \citenamefont {Polkovnikov},\ and\
  \citenamefont {Rigol}}]{Rigol_AIP_2016}%
  \BibitemOpen
  \bibfield  {author} {\bibinfo {author} {\bibfnamefont {Luca}\ \bibnamefont
  {D'Alessio}}, \bibinfo {author} {\bibfnamefont {Yariv}\ \bibnamefont
  {Kafri}}, \bibinfo {author} {\bibfnamefont {Anatoli}\ \bibnamefont
  {Polkovnikov}}, \ and\ \bibinfo {author} {\bibfnamefont {Marcos}\
  \bibnamefont {Rigol}},\ }\bibfield  {title} {\enquote {\bibinfo {title} {From
  quantum chaos and eigenstate thermalization to statistical mechanics and
  thermodynamics},}\ }\href {\doibase 10.1080/00018732.2016.1198134} {\bibfield
   {journal} {\bibinfo  {journal} {Advances in Physics}\ }\textbf {\bibinfo
  {volume} {65}},\ \bibinfo {pages} {239--362} (\bibinfo {year} {2016})},\
  \Eprint {http://arxiv.org/abs/https://doi.org/10.1080/00018732.2016.1198134}
  {https://doi.org/10.1080/00018732.2016.1198134} \BibitemShut {NoStop}%
\bibitem [{\citenamefont {Borgonovi}\ \emph {et~al.}(2016)\citenamefont
  {Borgonovi}, \citenamefont {Izrailev}, \citenamefont {Santos},\ and\
  \citenamefont {Zelevinsky}}]{BORGONOVI20161}%
  \BibitemOpen
  \bibfield  {author} {\bibinfo {author} {\bibfnamefont {F.}~\bibnamefont
  {Borgonovi}}, \bibinfo {author} {\bibfnamefont {F.M.}\ \bibnamefont
  {Izrailev}}, \bibinfo {author} {\bibfnamefont {L.F.}\ \bibnamefont {Santos}},
  \ and\ \bibinfo {author} {\bibfnamefont {V.G.}\ \bibnamefont {Zelevinsky}},\
  }\bibfield  {title} {\enquote {\bibinfo {title} {Quantum chaos and
  thermalization in isolated systems of interacting particles},}\ }\href
  {\doibase https://doi.org/10.1016/j.physrep.2016.02.005} {\bibfield
  {journal} {\bibinfo  {journal} {Physics Reports}\ }\textbf {\bibinfo {volume}
  {626}},\ \bibinfo {pages} {1--58} (\bibinfo {year} {2016})},\ \bibinfo {note}
  {quantum chaos and thermalization in isolated systems of interacting
  particles}\BibitemShut {NoStop}%
\bibitem [{\citenamefont {Das}\ \emph {et~al.}(2018)\citenamefont {Das},
  \citenamefont {Chakrabarty}, \citenamefont {Dhar}, \citenamefont {Kundu},
  \citenamefont {Huse}, \citenamefont {Moessner}, \citenamefont {Ray},\ and\
  \citenamefont {Bhattacharjee}}]{Das_PRL_2018}%
  \BibitemOpen
  \bibfield  {author} {\bibinfo {author} {\bibfnamefont {Avijit}\ \bibnamefont
  {Das}}, \bibinfo {author} {\bibfnamefont {Saurish}\ \bibnamefont
  {Chakrabarty}}, \bibinfo {author} {\bibfnamefont {Abhishek}\ \bibnamefont
  {Dhar}}, \bibinfo {author} {\bibfnamefont {Anupam}\ \bibnamefont {Kundu}},
  \bibinfo {author} {\bibfnamefont {David~A.}\ \bibnamefont {Huse}}, \bibinfo
  {author} {\bibfnamefont {Roderich}\ \bibnamefont {Moessner}}, \bibinfo
  {author} {\bibfnamefont {Samriddhi~Sankar}\ \bibnamefont {Ray}}, \ and\
  \bibinfo {author} {\bibfnamefont {Subhro}\ \bibnamefont {Bhattacharjee}},\
  }\bibfield  {title} {\enquote {\bibinfo {title} {Light-cone spreading of
  perturbations and the butterfly effect in a classical spin chain},}\ }\href
  {\doibase 10.1103/PhysRevLett.121.024101} {\bibfield  {journal} {\bibinfo
  {journal} {Phys. Rev. Lett.}\ }\textbf {\bibinfo {volume} {121}},\ \bibinfo
  {pages} {024101} (\bibinfo {year} {2018})}\BibitemShut {NoStop}%
\bibitem [{\citenamefont {Pollack}\ \emph {et~al.}(2020)\citenamefont
  {Pollack}, \citenamefont {Rozali}, \citenamefont {Sully},\ and\ \citenamefont
  {Wakeham}}]{PollackPRL2020}%
  \BibitemOpen
  \bibfield  {author} {\bibinfo {author} {\bibfnamefont {Jason}\ \bibnamefont
  {Pollack}}, \bibinfo {author} {\bibfnamefont {Moshe}\ \bibnamefont {Rozali}},
  \bibinfo {author} {\bibfnamefont {James}\ \bibnamefont {Sully}}, \ and\
  \bibinfo {author} {\bibfnamefont {David}\ \bibnamefont {Wakeham}},\
  }\bibfield  {title} {\enquote {\bibinfo {title} {Eigenstate thermalization
  and disorder averaging in gravity},}\ }\href {\doibase
  10.1103/PhysRevLett.125.021601} {\bibfield  {journal} {\bibinfo  {journal}
  {Phys. Rev. Lett.}\ }\textbf {\bibinfo {volume} {125}},\ \bibinfo {pages}
  {021601} (\bibinfo {year} {2020})}\BibitemShut {NoStop}%
\bibitem [{\citenamefont {Murugan}\ \emph {et~al.}(2021)\citenamefont
  {Murugan}, \citenamefont {Kumar}, \citenamefont {Bhattacharjee},\ and\
  \citenamefont {Ray}}]{Murugan_PRL_2021}%
  \BibitemOpen
  \bibfield  {author} {\bibinfo {author} {\bibfnamefont {Sugan~Durai}\
  \bibnamefont {Murugan}}, \bibinfo {author} {\bibfnamefont {Dheeraj}\
  \bibnamefont {Kumar}}, \bibinfo {author} {\bibfnamefont {Subhro}\
  \bibnamefont {Bhattacharjee}}, \ and\ \bibinfo {author} {\bibfnamefont
  {Samriddhi~Sankar}\ \bibnamefont {Ray}},\ }\bibfield  {title} {\enquote
  {\bibinfo {title} {Many-body chaos in thermalized fluids},}\ }\href {\doibase
  10.1103/PhysRevLett.127.124501} {\bibfield  {journal} {\bibinfo  {journal}
  {Phys. Rev. Lett.}\ }\textbf {\bibinfo {volume} {127}},\ \bibinfo {pages}
  {124501} (\bibinfo {year} {2021})}\BibitemShut {NoStop}%
\bibitem [{\citenamefont {Cotler}\ \emph
  {et~al.}(2017{\natexlab{a}})\citenamefont {Cotler}, \citenamefont {Gur-Ari},
  \citenamefont {Hanada}, \citenamefont {Polchinski}, \citenamefont {Saad},
  \citenamefont {Shenker}, \citenamefont {Stanford}, \citenamefont
  {Streicher},\ and\ \citenamefont {Tezuka}}]{polchinski2016}%
  \BibitemOpen
  \bibfield  {author} {\bibinfo {author} {\bibfnamefont {Jordan~S.}\
  \bibnamefont {Cotler}}, \bibinfo {author} {\bibfnamefont {Guy}\ \bibnamefont
  {Gur-Ari}}, \bibinfo {author} {\bibfnamefont {Masanori}\ \bibnamefont
  {Hanada}}, \bibinfo {author} {\bibfnamefont {Joseph}\ \bibnamefont
  {Polchinski}}, \bibinfo {author} {\bibfnamefont {Phil}\ \bibnamefont {Saad}},
  \bibinfo {author} {\bibfnamefont {Stephen~H.}\ \bibnamefont {Shenker}},
  \bibinfo {author} {\bibfnamefont {Douglas}\ \bibnamefont {Stanford}},
  \bibinfo {author} {\bibfnamefont {Alexandre}\ \bibnamefont {Streicher}}, \
  and\ \bibinfo {author} {\bibfnamefont {Masaki}\ \bibnamefont {Tezuka}},\
  }\bibfield  {title} {\enquote {\bibinfo {title} {{Black Holes and Random
  Matrices}},}\ }\href {\doibase 10.1007/JHEP05(2017)118} {\bibfield  {journal}
  {\bibinfo  {journal} {JHEP}\ }\textbf {\bibinfo {volume} {05}},\ \bibinfo
  {pages} {118} (\bibinfo {year} {2017}{\natexlab{a}})},\ \bibinfo {note}
  {[Erratum: JHEP 09, 002 (2018)]},\ \Eprint {http://arxiv.org/abs/1611.04650}
  {arXiv:1611.04650 [hep-th]} \BibitemShut {NoStop}%
\bibitem [{\citenamefont {Saad}\ \emph {et~al.}(2018)\citenamefont {Saad},
  \citenamefont {Shenker},\ and\ \citenamefont {Stanford}}]{sss2}%
  \BibitemOpen
  \bibfield  {author} {\bibinfo {author} {\bibfnamefont {Phil}\ \bibnamefont
  {Saad}}, \bibinfo {author} {\bibfnamefont {Stephen~H.}\ \bibnamefont
  {Shenker}}, \ and\ \bibinfo {author} {\bibfnamefont {Douglas}\ \bibnamefont
  {Stanford}},\ }\bibfield  {title} {\enquote {\bibinfo {title} {{A
  semiclassical ramp in SYK and in gravity}},}\ }\href@noop {} {\  (\bibinfo
  {year} {2018})},\ \Eprint {http://arxiv.org/abs/1806.06840} {arXiv:1806.06840
  [hep-th]} \BibitemShut {NoStop}%
\bibitem [{\citenamefont {Dyer}\ and\ \citenamefont
  {Gur-Ari}(2017)}]{Dyer:2016pou}%
  \BibitemOpen
  \bibfield  {author} {\bibinfo {author} {\bibfnamefont {Ethan}\ \bibnamefont
  {Dyer}}\ and\ \bibinfo {author} {\bibfnamefont {Guy}\ \bibnamefont
  {Gur-Ari}},\ }\bibfield  {title} {\enquote {\bibinfo {title} {{2D CFT
  Partition Functions at Late Times}},}\ }\href {\doibase
  10.1007/JHEP08(2017)075} {\bibfield  {journal} {\bibinfo  {journal} {JHEP}\
  }\textbf {\bibinfo {volume} {08}},\ \bibinfo {pages} {075} (\bibinfo {year}
  {2017})},\ \Eprint {http://arxiv.org/abs/1611.04592} {arXiv:1611.04592
  [hep-th]} \BibitemShut {NoStop}%
\bibitem [{\citenamefont {Nivedita}\ \emph {et~al.}(2020)\citenamefont
  {Nivedita}, \citenamefont {Shackleton},\ and\ \citenamefont
  {Sachdev}}]{Nivedita_PRE_2020}%
  \BibitemOpen
  \bibfield  {author} {\bibinfo {author} {\bibnamefont {Nivedita}}, \bibinfo
  {author} {\bibfnamefont {Henry}\ \bibnamefont {Shackleton}}, \ and\ \bibinfo
  {author} {\bibfnamefont {Subir}\ \bibnamefont {Sachdev}},\ }\bibfield
  {title} {\enquote {\bibinfo {title} {Spectral form factors of clean and
  random quantum ising chains},}\ }\href {\doibase 10.1103/PhysRevE.101.042136}
  {\bibfield  {journal} {\bibinfo  {journal} {Phys. Rev. E}\ }\textbf {\bibinfo
  {volume} {101}},\ \bibinfo {pages} {042136} (\bibinfo {year}
  {2020})}\BibitemShut {NoStop}%
\bibitem [{\citenamefont {del Campo}\ \emph {et~al.}(2017)\citenamefont {del
  Campo}, \citenamefont {Molina-Vilaplana},\ and\ \citenamefont
  {Sonner}}]{sonner2017}%
  \BibitemOpen
  \bibfield  {author} {\bibinfo {author} {\bibfnamefont {A.}~\bibnamefont {del
  Campo}}, \bibinfo {author} {\bibfnamefont {J.}~\bibnamefont
  {Molina-Vilaplana}}, \ and\ \bibinfo {author} {\bibfnamefont
  {J.}~\bibnamefont {Sonner}},\ }\bibfield  {title} {\enquote {\bibinfo {title}
  {{Scrambling the spectral form factor: unitarity constraints and exact
  results}},}\ }\href {\doibase 10.1103/PhysRevD.95.126008} {\bibfield
  {journal} {\bibinfo  {journal} {Phys. Rev. D}\ }\textbf {\bibinfo {volume}
  {95}},\ \bibinfo {pages} {126008} (\bibinfo {year} {2017})},\ \Eprint
  {http://arxiv.org/abs/1702.04350} {arXiv:1702.04350 [hep-th]} \BibitemShut
  {NoStop}%
\bibitem [{\citenamefont {Bena}\ \emph {et~al.}(2005)\citenamefont {Bena},
  \citenamefont {Droz},\ and\ \citenamefont {Lipkowski}}]{2005}%
  \BibitemOpen
  \bibfield  {author} {\bibinfo {author} {\bibfnamefont {I}~\bibnamefont
  {Bena}}, \bibinfo {author} {\bibfnamefont {M}~\bibnamefont {Droz}}, \ and\
  \bibinfo {author} {\bibfnamefont {A}~\bibnamefont {Lipkowski}},\ }\bibfield
  {title} {\enquote {\bibinfo {title} {Statistical mechanics of equilibrium and
  nonequilibrium phase transitions: The yang-lee formalism},}\ }\href {\doibase
  10.1142/s0217979205032759} {\bibfield  {journal} {\bibinfo  {journal}
  {International Journal of Modern Physics B}\ }\textbf {\bibinfo {volume}
  {19}},\ \bibinfo {pages} {4269–4329} (\bibinfo {year} {2005})}\BibitemShut
  {NoStop}%
\bibitem [{\citenamefont {Harrow}\ \emph {et~al.}(2020)\citenamefont {Harrow},
  \citenamefont {Mehraban},\ and\ \citenamefont {Soleimanifar}}]{2020}%
  \BibitemOpen
  \bibfield  {author} {\bibinfo {author} {\bibfnamefont {Aram~W.}\ \bibnamefont
  {Harrow}}, \bibinfo {author} {\bibfnamefont {Saeed}\ \bibnamefont
  {Mehraban}}, \ and\ \bibinfo {author} {\bibfnamefont {Mehdi}\ \bibnamefont
  {Soleimanifar}},\ }\bibfield  {title} {\enquote {\bibinfo {title} {Classical
  algorithms, correlation decay, and complex zeros of partition functions of
  quantum many-body systems},}\ }\href {\doibase 10.1145/3357713.3384322}
  {\bibfield  {journal} {\bibinfo  {journal} {Proceedings of the 52nd Annual
  ACM SIGACT Symposium on Theory of Computing}\ } (\bibinfo {year} {2020}),\
  10.1145/3357713.3384322}\BibitemShut {NoStop}%
\bibitem [{\citenamefont {Cotler}\ \emph
  {et~al.}(2017{\natexlab{b}})\citenamefont {Cotler}, \citenamefont
  {Hunter-Jones}, \citenamefont {Liu},\ and\ \citenamefont
  {Yoshida}}]{cotler2017}%
  \BibitemOpen
  \bibfield  {author} {\bibinfo {author} {\bibfnamefont {Jordan}\ \bibnamefont
  {Cotler}}, \bibinfo {author} {\bibfnamefont {Nicholas}\ \bibnamefont
  {Hunter-Jones}}, \bibinfo {author} {\bibfnamefont {Junyu}\ \bibnamefont
  {Liu}}, \ and\ \bibinfo {author} {\bibfnamefont {Beni}\ \bibnamefont
  {Yoshida}},\ }\bibfield  {title} {\enquote {\bibinfo {title} {{Chaos,
  Complexity, and Random Matrices}},}\ }\href {\doibase
  10.1007/JHEP11(2017)048} {\bibfield  {journal} {\bibinfo  {journal} {JHEP}\
  }\textbf {\bibinfo {volume} {11}},\ \bibinfo {pages} {048} (\bibinfo {year}
  {2017}{\natexlab{b}})},\ \Eprint {http://arxiv.org/abs/1706.05400}
  {arXiv:1706.05400 [hep-th]} \BibitemShut {NoStop}%
\bibitem [{\citenamefont {Liu}(2018)}]{liu2018}%
  \BibitemOpen
  \bibfield  {author} {\bibinfo {author} {\bibfnamefont {Junyu}\ \bibnamefont
  {Liu}},\ }\bibfield  {title} {\enquote {\bibinfo {title} {{Spectral form
  factors and late time quantum chaos}},}\ }\href {\doibase
  10.1103/PhysRevD.98.086026} {\bibfield  {journal} {\bibinfo  {journal} {Phys.
  Rev. D}\ }\textbf {\bibinfo {volume} {98}},\ \bibinfo {pages} {086026}
  (\bibinfo {year} {2018})},\ \Eprint {http://arxiv.org/abs/1806.05316}
  {arXiv:1806.05316 [hep-th]} \BibitemShut {NoStop}%
\bibitem [{\citenamefont {Winer}\ \emph {et~al.}(2020)\citenamefont {Winer},
  \citenamefont {Jian},\ and\ \citenamefont {Swingle}}]{Winer:2020mdc}%
  \BibitemOpen
  \bibfield  {author} {\bibinfo {author} {\bibfnamefont {Michael}\ \bibnamefont
  {Winer}}, \bibinfo {author} {\bibfnamefont {Shao-Kai}\ \bibnamefont {Jian}},
  \ and\ \bibinfo {author} {\bibfnamefont {Brian}\ \bibnamefont {Swingle}},\
  }\bibfield  {title} {\enquote {\bibinfo {title} {{An exponential ramp in the
  quadratic Sachdev-Ye-Kitaev model}},}\ }\href {\doibase
  10.1103/PhysRevLett.125.250602} {\bibfield  {journal} {\bibinfo  {journal}
  {Phys. Rev. Lett.}\ }\textbf {\bibinfo {volume} {125}},\ \bibinfo {pages}
  {250602} (\bibinfo {year} {2020})},\ \Eprint
  {http://arxiv.org/abs/2006.15152} {arXiv:2006.15152 [cond-mat.stat-mech]}
  \BibitemShut {NoStop}%
\bibitem [{\citenamefont {Liao}\ \emph {et~al.}(2020)\citenamefont {Liao},
  \citenamefont {Vikram},\ and\ \citenamefont {Galitski}}]{Liao:2020lac}%
  \BibitemOpen
  \bibfield  {author} {\bibinfo {author} {\bibfnamefont {Yunxiang}\
  \bibnamefont {Liao}}, \bibinfo {author} {\bibfnamefont {Amit}\ \bibnamefont
  {Vikram}}, \ and\ \bibinfo {author} {\bibfnamefont {Victor}\ \bibnamefont
  {Galitski}},\ }\bibfield  {title} {\enquote {\bibinfo {title} {{Many-body
  level statistics of single-particle quantum chaos}},}\ }\href {\doibase
  10.1103/PhysRevLett.125.250601} {\bibfield  {journal} {\bibinfo  {journal}
  {Phys. Rev. Lett.}\ }\textbf {\bibinfo {volume} {125}},\ \bibinfo {pages}
  {250601} (\bibinfo {year} {2020})},\ \Eprint
  {http://arxiv.org/abs/2005.08991} {arXiv:2005.08991 [cond-mat.stat-mech]}
  \BibitemShut {NoStop}%
\bibitem [{\citenamefont {Kudler-Flam}\ \emph {et~al.}(2020)\citenamefont
  {Kudler-Flam}, \citenamefont {Nie},\ and\ \citenamefont
  {Ryu}}]{Kudler-Flam2019JHEP}%
  \BibitemOpen
  \bibfield  {author} {\bibinfo {author} {\bibfnamefont {Jonah}\ \bibnamefont
  {Kudler-Flam}}, \bibinfo {author} {\bibfnamefont {Laimei}\ \bibnamefont
  {Nie}}, \ and\ \bibinfo {author} {\bibfnamefont {Shinsei}\ \bibnamefont
  {Ryu}},\ }\bibfield  {title} {\enquote {\bibinfo {title} {{Conformal field
  theory and the web of quantum chaos diagnostics}},}\ }\href {\doibase
  10.1007/JHEP01(2020)175} {\bibfield  {journal} {\bibinfo  {journal} {JHEP}\
  }\textbf {\bibinfo {volume} {01}},\ \bibinfo {pages} {175} (\bibinfo {year}
  {2020})},\ \Eprint {http://arxiv.org/abs/1910.14575} {arXiv:1910.14575
  [hep-th]} \BibitemShut {NoStop}%
\bibitem [{\citenamefont {Ludwig}(2016)}]{Ludwig_PS_2015}%
  \BibitemOpen
  \bibfield  {author} {\bibinfo {author} {\bibfnamefont {Andreas W~W}\
  \bibnamefont {Ludwig}},\ }\bibfield  {title} {\enquote {\bibinfo {title}
  {Topological phases: classification of topological insulators and
  superconductors of non-interacting fermions, and beyond},}\ }\href
  {http://stacks.iop.org/1402-4896/2016/i=T168/a=014001} {\bibfield  {journal}
  {\bibinfo  {journal} {Physica Scripta}\ }\textbf {\bibinfo {volume} {2016}},\
  \bibinfo {pages} {014001} (\bibinfo {year} {2016})}\BibitemShut {NoStop}%
\bibitem [{\citenamefont {Chiu}\ \emph {et~al.}(2016)\citenamefont {Chiu},
  \citenamefont {Teo}, \citenamefont {Schnyder},\ and\ \citenamefont
  {Ryu}}]{Chiu_RMP_2016}%
  \BibitemOpen
  \bibfield  {author} {\bibinfo {author} {\bibfnamefont {Ching-Kai}\
  \bibnamefont {Chiu}}, \bibinfo {author} {\bibfnamefont {Jeffrey C.~Y.}\
  \bibnamefont {Teo}}, \bibinfo {author} {\bibfnamefont {Andreas~P.}\
  \bibnamefont {Schnyder}}, \ and\ \bibinfo {author} {\bibfnamefont {Shinsei}\
  \bibnamefont {Ryu}},\ }\bibfield  {title} {\enquote {\bibinfo {title}
  {Classification of topological quantum matter with symmetries},}\ }\href
  {\doibase 10.1103/RevModPhys.88.035005} {\bibfield  {journal} {\bibinfo
  {journal} {Rev. Mod. Phys.}\ }\textbf {\bibinfo {volume} {88}},\ \bibinfo
  {pages} {035005} (\bibinfo {year} {2016})}\BibitemShut {NoStop}%
\bibitem [{\citenamefont {Hasan}\ and\ \citenamefont
  {Kane}(2010)}]{Hasan_RMP_2010}%
  \BibitemOpen
  \bibfield  {author} {\bibinfo {author} {\bibfnamefont {M.~Z.}\ \bibnamefont
  {Hasan}}\ and\ \bibinfo {author} {\bibfnamefont {C.~L.}\ \bibnamefont
  {Kane}},\ }\bibfield  {title} {\enquote {\bibinfo {title}
  {\textit{Colloquium} : Topological insulators},}\ }\href {\doibase
  10.1103/RevModPhys.82.3045} {\bibfield  {journal} {\bibinfo  {journal} {Rev.
  Mod. Phys.}\ }\textbf {\bibinfo {volume} {82}},\ \bibinfo {pages}
  {3045--3067} (\bibinfo {year} {2010})}\BibitemShut {NoStop}%
\bibitem [{\citenamefont {Qi}\ and\ \citenamefont {Zhang}(2011)}]{Qi_RMP_2011}%
  \BibitemOpen
  \bibfield  {author} {\bibinfo {author} {\bibfnamefont {Xiao-Liang}\
  \bibnamefont {Qi}}\ and\ \bibinfo {author} {\bibfnamefont {Shou-Cheng}\
  \bibnamefont {Zhang}},\ }\bibfield  {title} {\enquote {\bibinfo {title}
  {Topological insulators and superconductors},}\ }\href {\doibase
  10.1103/RevModPhys.83.1057} {\bibfield  {journal} {\bibinfo  {journal} {Rev.
  Mod. Phys.}\ }\textbf {\bibinfo {volume} {83}},\ \bibinfo {pages}
  {1057--1110} (\bibinfo {year} {2011})}\BibitemShut {NoStop}%
\bibitem [{\citenamefont {Asb{\'o}th}\ \emph {et~al.}(2016)\citenamefont
  {Asb{\'o}th}, \citenamefont {Oroszl{\'a}ny},\ and\ \citenamefont
  {P{\'a}lyi}}]{asboth2016short}%
  \BibitemOpen
  \bibfield  {author} {\bibinfo {author} {\bibfnamefont {J{\'a}nos~K}\
  \bibnamefont {Asb{\'o}th}}, \bibinfo {author} {\bibfnamefont
  {L{\'a}szl{\'o}}\ \bibnamefont {Oroszl{\'a}ny}}, \ and\ \bibinfo {author}
  {\bibfnamefont {Andr{\'a}s}\ \bibnamefont {P{\'a}lyi}},\ }\bibfield  {title}
  {\enquote {\bibinfo {title} {A short course on topological insulators},}\
  }\href@noop {} {\bibfield  {journal} {\bibinfo  {journal} {Lecture notes in
  physics}\ }\textbf {\bibinfo {volume} {919}},\ \bibinfo {pages} {166}
  (\bibinfo {year} {2016})}\BibitemShut {NoStop}%
\bibitem [{\citenamefont {Su}\ \emph {et~al.}(1979)\citenamefont {Su},
  \citenamefont {Schrieffer},\ and\ \citenamefont {Heeger}}]{ssh}%
  \BibitemOpen
  \bibfield  {author} {\bibinfo {author} {\bibfnamefont {W.~P.}\ \bibnamefont
  {Su}}, \bibinfo {author} {\bibfnamefont {J.~R.}\ \bibnamefont {Schrieffer}},
  \ and\ \bibinfo {author} {\bibfnamefont {A.~J.}\ \bibnamefont {Heeger}},\
  }\bibfield  {title} {\enquote {\bibinfo {title} {Solitons in
  polyacetylene},}\ }\href {\doibase 10.1103/PhysRevLett.42.1698} {\bibfield
  {journal} {\bibinfo  {journal} {Phys. Rev. Lett.}\ }\textbf {\bibinfo
  {volume} {42}},\ \bibinfo {pages} {1698--1701} (\bibinfo {year}
  {1979})}\BibitemShut {NoStop}%
\bibitem [{\citenamefont {Benalcazar}\ \emph {et~al.}(2017)\citenamefont
  {Benalcazar}, \citenamefont {Bernevig},\ and\ \citenamefont
  {Hughes}}]{benalcazar2017quantized}%
  \BibitemOpen
  \bibfield  {author} {\bibinfo {author} {\bibfnamefont {Wladimir~A}\
  \bibnamefont {Benalcazar}}, \bibinfo {author} {\bibfnamefont {B~Andrei}\
  \bibnamefont {Bernevig}}, \ and\ \bibinfo {author} {\bibfnamefont {Taylor~L}\
  \bibnamefont {Hughes}},\ }\bibfield  {title} {\enquote {\bibinfo {title}
  {Quantized electric multipole insulators},}\ }\href@noop {} {\bibfield
  {journal} {\bibinfo  {journal} {Science}\ }\textbf {\bibinfo {volume}
  {357}},\ \bibinfo {pages} {61--66} (\bibinfo {year} {2017})}\BibitemShut
  {NoStop}%
\bibitem [{\citenamefont {Chen}(2022)}]{Chen:2022hbi}%
  \BibitemOpen
  \bibfield  {author} {\bibinfo {author} {\bibfnamefont {Yiming}\ \bibnamefont
  {Chen}},\ }\bibfield  {title} {\enquote {\bibinfo {title} {{Spectral form
  factor for free large N gauge theory and strings}},}\ }\href {\doibase
  10.1007/JHEP06(2022)137} {\bibfield  {journal} {\bibinfo  {journal} {JHEP}\
  }\textbf {\bibinfo {volume} {06}},\ \bibinfo {pages} {137} (\bibinfo {year}
  {2022})},\ \Eprint {http://arxiv.org/abs/2202.04741} {arXiv:2202.04741
  [hep-th]} \BibitemShut {NoStop}%
\bibitem [{\citenamefont {Prange}(1997)}]{Prange}%
  \BibitemOpen
  \bibfield  {author} {\bibinfo {author} {\bibfnamefont {R.~E.}\ \bibnamefont
  {Prange}},\ }\bibfield  {title} {\enquote {\bibinfo {title} {The spectral
  form factor is not self-averaging},}\ }\href {\doibase
  10.1103/PhysRevLett.78.2280} {\bibfield  {journal} {\bibinfo  {journal}
  {Phys. Rev. Lett.}\ }\textbf {\bibinfo {volume} {78}},\ \bibinfo {pages}
  {2280--2283} (\bibinfo {year} {1997})}\BibitemShut {NoStop}%
\bibitem [{\citenamefont {Brézin}\ and\ \citenamefont
  {Hikami}(1997)}]{Br_zin_1997}%
  \BibitemOpen
  \bibfield  {author} {\bibinfo {author} {\bibfnamefont {E.}~\bibnamefont
  {Brézin}}\ and\ \bibinfo {author} {\bibfnamefont {S.}~\bibnamefont
  {Hikami}},\ }\bibfield  {title} {\enquote {\bibinfo {title} {Spectral form
  factor in a random matrix theory},}\ }\href {\doibase
  10.1103/physreve.55.4067} {\bibfield  {journal} {\bibinfo  {journal}
  {Physical Review E}\ }\textbf {\bibinfo {volume} {55}},\ \bibinfo {pages}
  {4067–4083} (\bibinfo {year} {1997})}\BibitemShut {NoStop}%
\bibitem [{\citenamefont {Cipolloni}\ \emph {et~al.}(2023)\citenamefont
  {Cipolloni}, \citenamefont {Erd{\H o}s},\ and\ \citenamefont
  {Schr{\"o}der}}]{Cipolloni2023}%
  \BibitemOpen
  \bibfield  {author} {\bibinfo {author} {\bibfnamefont {Giorgio}\ \bibnamefont
  {Cipolloni}}, \bibinfo {author} {\bibfnamefont {L{\'a}szl{\'o}}\ \bibnamefont
  {Erd{\H o}s}}, \ and\ \bibinfo {author} {\bibfnamefont {Dominik}\
  \bibnamefont {Schr{\"o}der}},\ }\bibfield  {title} {\enquote {\bibinfo
  {title} {On the spectral form factor for random matrices},}\ }\href {\doibase
  10.1007/s00220-023-04692-y} {\bibfield  {journal} {\bibinfo  {journal}
  {Communications in Mathematical Physics}\ }\textbf {\bibinfo {volume}
  {401}},\ \bibinfo {pages} {1665--1700} (\bibinfo {year} {2023})}\BibitemShut
  {NoStop}%
\bibitem [{\citenamefont {Sierant}\ \emph {et~al.}(2020)\citenamefont
  {Sierant}, \citenamefont {Delande},\ and\ \citenamefont
  {Zakrzewski}}]{SierantPRL2020}%
  \BibitemOpen
  \bibfield  {author} {\bibinfo {author} {\bibfnamefont {Piotr}\ \bibnamefont
  {Sierant}}, \bibinfo {author} {\bibfnamefont {Dominique}\ \bibnamefont
  {Delande}}, \ and\ \bibinfo {author} {\bibfnamefont {Jakub}\ \bibnamefont
  {Zakrzewski}},\ }\bibfield  {title} {\enquote {\bibinfo {title} {Thouless
  time analysis of anderson and many-body localization transitions},}\ }\href
  {\doibase 10.1103/PhysRevLett.124.186601} {\bibfield  {journal} {\bibinfo
  {journal} {Phys. Rev. Lett.}\ }\textbf {\bibinfo {volume} {124}},\ \bibinfo
  {pages} {186601} (\bibinfo {year} {2020})}\BibitemShut {NoStop}%
\bibitem [{\citenamefont {Brezin}\ and\ \citenamefont
  {Zee}(1993)}]{Brezin:1993qg}%
  \BibitemOpen
  \bibfield  {author} {\bibinfo {author} {\bibfnamefont {E.}~\bibnamefont
  {Brezin}}\ and\ \bibinfo {author} {\bibfnamefont {A.}~\bibnamefont {Zee}},\
  }\bibfield  {title} {\enquote {\bibinfo {title} {{Universality of the
  correlations between eigenvalues of large random matrices}},}\ }\href
  {\doibase 10.1016/0550-3213(93)90121-5} {\bibfield  {journal} {\bibinfo
  {journal} {Nucl. Phys. B}\ }\textbf {\bibinfo {volume} {402}},\ \bibinfo
  {pages} {613--627} (\bibinfo {year} {1993})}\BibitemShut {NoStop}%
\bibitem [{\citenamefont {Mehta}(2004)}]{Meh2004}%
  \BibitemOpen
  \bibfield  {author} {\bibinfo {author} {\bibfnamefont {Madan~Lal}\
  \bibnamefont {Mehta}},\ }\href@noop {} {\emph {\bibinfo {title} {Random
  Matrices}}},\ \bibinfo {edition} {3rd}\ ed.\ (\bibinfo {year}
  {2004})\BibitemShut {NoStop}%
\bibitem [{\citenamefont {Cotler}\ \emph
  {et~al.}(2017{\natexlab{c}})\citenamefont {Cotler}, \citenamefont
  {Hunter-Jones}, \citenamefont {Liu},\ and\ \citenamefont
  {Yoshida}}]{Cotler:2017jue}%
  \BibitemOpen
  \bibfield  {author} {\bibinfo {author} {\bibfnamefont {Jordan}\ \bibnamefont
  {Cotler}}, \bibinfo {author} {\bibfnamefont {Nicholas}\ \bibnamefont
  {Hunter-Jones}}, \bibinfo {author} {\bibfnamefont {Junyu}\ \bibnamefont
  {Liu}}, \ and\ \bibinfo {author} {\bibfnamefont {Beni}\ \bibnamefont
  {Yoshida}},\ }\bibfield  {title} {\enquote {\bibinfo {title} {{Chaos,
  Complexity, and Random Matrices}},}\ }\href {\doibase
  10.1007/JHEP11(2017)048} {\bibfield  {journal} {\bibinfo  {journal} {JHEP}\
  }\textbf {\bibinfo {volume} {11}},\ \bibinfo {pages} {048} (\bibinfo {year}
  {2017}{\natexlab{c}})},\ \Eprint {http://arxiv.org/abs/1706.05400}
  {arXiv:1706.05400 [hep-th]} \BibitemShut {NoStop}%
\bibitem [{\citenamefont {Altland}\ and\ \citenamefont
  {Zirnbauer}(1997)}]{Altland_PRB_1997}%
  \BibitemOpen
  \bibfield  {author} {\bibinfo {author} {\bibfnamefont {Alexander}\
  \bibnamefont {Altland}}\ and\ \bibinfo {author} {\bibfnamefont {Martin~R.}\
  \bibnamefont {Zirnbauer}},\ }\bibfield  {title} {\enquote {\bibinfo {title}
  {Nonstandard symmetry classes in mesoscopic normal-superconducting hybrid
  structures},}\ }\href {\doibase 10.1103/PhysRevB.55.1142} {\bibfield
  {journal} {\bibinfo  {journal} {Phys. Rev. B}\ }\textbf {\bibinfo {volume}
  {55}},\ \bibinfo {pages} {1142--1161} (\bibinfo {year} {1997})}\BibitemShut
  {NoStop}%
\bibitem [{\citenamefont {Agarwala}\ \emph {et~al.}(2017)\citenamefont
  {Agarwala}, \citenamefont {Haldar},\ and\ \citenamefont
  {Shenoy}}]{Agarwala_AOP_2017}%
  \BibitemOpen
  \bibfield  {author} {\bibinfo {author} {\bibfnamefont {Adhip}\ \bibnamefont
  {Agarwala}}, \bibinfo {author} {\bibfnamefont {Arijit}\ \bibnamefont
  {Haldar}}, \ and\ \bibinfo {author} {\bibfnamefont {Vijay~B.}\ \bibnamefont
  {Shenoy}},\ }\bibfield  {title} {\enquote {\bibinfo {title} {The tenfold way
  redux: Fermionic systems with n-body interactions},}\ }\href {\doibase
  https://doi.org/10.1016/j.aop.2017.07.016} {\bibfield  {journal} {\bibinfo
  {journal} {Annals of Physics}\ }\textbf {\bibinfo {volume} {385}},\ \bibinfo
  {pages} {469 -- 511} (\bibinfo {year} {2017})}\BibitemShut {NoStop}%
\bibitem [{\citenamefont {Sakurai}(1993)}]{Sakurai1993Modern}%
  \BibitemOpen
  \bibfield  {author} {\bibinfo {author} {\bibfnamefont {J.~J.}\ \bibnamefont
  {Sakurai}},\ }\href {http://www.worldcat.org/isbn/0201539292} {\emph
  {\bibinfo {title} {Modern Quantum Mechanics (Revised Edition)}}},\ \bibinfo
  {edition} {1st}\ ed.\ (\bibinfo  {publisher} {Addison Wesley},\ \bibinfo
  {year} {1993})\BibitemShut {NoStop}%
\bibitem [{\citenamefont {Griffiths}\ and\ \citenamefont
  {Schroeter}(2018)}]{griffiths2018introduction}%
  \BibitemOpen
  \bibfield  {author} {\bibinfo {author} {\bibfnamefont {David~J}\ \bibnamefont
  {Griffiths}}\ and\ \bibinfo {author} {\bibfnamefont {Darrell~F}\ \bibnamefont
  {Schroeter}},\ }\href@noop {} {\emph {\bibinfo {title} {Introduction to
  quantum mechanics}}}\ (\bibinfo  {publisher} {Cambridge university press},\
  \bibinfo {year} {2018})\BibitemShut {NoStop}%
\bibitem [{\citenamefont {Kang}\ \emph {et~al.}(2019)\citenamefont {Kang},
  \citenamefont {Shiozaki},\ and\ \citenamefont {Cho}}]{Kang_prb_2019}%
  \BibitemOpen
  \bibfield  {author} {\bibinfo {author} {\bibfnamefont {Byungmin}\
  \bibnamefont {Kang}}, \bibinfo {author} {\bibfnamefont {Ken}\ \bibnamefont
  {Shiozaki}}, \ and\ \bibinfo {author} {\bibfnamefont {Gil~Young}\
  \bibnamefont {Cho}},\ }\bibfield  {title} {\enquote {\bibinfo {title}
  {Many-body order parameters for multipoles in solids},}\ }\href {\doibase
  10.1103/PhysRevB.100.245134} {\bibfield  {journal} {\bibinfo  {journal}
  {Phys. Rev. B}\ }\textbf {\bibinfo {volume} {100}},\ \bibinfo {pages}
  {245134} (\bibinfo {year} {2019})}\BibitemShut {NoStop}%
\bibitem [{\citenamefont {Dyson}(1962)}]{Dyson:1962es}%
  \BibitemOpen
  \bibfield  {author} {\bibinfo {author} {\bibfnamefont {F.~J.}\ \bibnamefont
  {Dyson}},\ }\bibfield  {title} {\enquote {\bibinfo {title} {{Statistical
  theory of the energy levels of complex systems. I}},}\ }\href {\doibase
  10.1063/1.1703773} {\bibfield  {journal} {\bibinfo  {journal} {J. Math.
  Phys.}\ }\textbf {\bibinfo {volume} {3}},\ \bibinfo {pages} {140--156}
  (\bibinfo {year} {1962})}\BibitemShut {NoStop}%
\bibitem [{\citenamefont {Bertini}\ \emph {et~al.}(2018)\citenamefont
  {Bertini}, \citenamefont {Kos},\ and\ \citenamefont {Prosen}}]{prosen}%
  \BibitemOpen
  \bibfield  {author} {\bibinfo {author} {\bibfnamefont {Bruno}\ \bibnamefont
  {Bertini}}, \bibinfo {author} {\bibfnamefont {Pavel}\ \bibnamefont {Kos}}, \
  and\ \bibinfo {author} {\bibfnamefont {Toma\ifmmode
  \check{z}\else~\v{z}\fi{}}\ \bibnamefont {Prosen}},\ }\bibfield  {title}
  {\enquote {\bibinfo {title} {Exact spectral form factor in a minimal model of
  many-body quantum chaos},}\ }\href {\doibase 10.1103/PhysRevLett.121.264101}
  {\bibfield  {journal} {\bibinfo  {journal} {Phys. Rev. Lett.}\ }\textbf
  {\bibinfo {volume} {121}},\ \bibinfo {pages} {264101} (\bibinfo {year}
  {2018})}\BibitemShut {NoStop}%
\bibitem [{\citenamefont {Chan}\ \emph {et~al.}(2018)\citenamefont {Chan},
  \citenamefont {De~Luca},\ and\ \citenamefont {Chalker}}]{chalker}%
  \BibitemOpen
  \bibfield  {author} {\bibinfo {author} {\bibfnamefont {Amos}\ \bibnamefont
  {Chan}}, \bibinfo {author} {\bibfnamefont {Andrea}\ \bibnamefont {De~Luca}},
  \ and\ \bibinfo {author} {\bibfnamefont {J.~T.}\ \bibnamefont {Chalker}},\
  }\bibfield  {title} {\enquote {\bibinfo {title} {Spectral statistics in
  spatially extended chaotic quantum many-body systems},}\ }\href {\doibase
  10.1103/PhysRevLett.121.060601} {\bibfield  {journal} {\bibinfo  {journal}
  {Phys. Rev. Lett.}\ }\textbf {\bibinfo {volume} {121}},\ \bibinfo {pages}
  {060601} (\bibinfo {year} {2018})}\BibitemShut {NoStop}%
\bibitem [{\citenamefont {Tezuka}\ \emph {et~al.}(2023)\citenamefont {Tezuka},
  \citenamefont {Oktay}, \citenamefont {Rinaldi}, \citenamefont {Hanada},\ and\
  \citenamefont {Nori}}]{Tezuka_PRB_2023}%
  \BibitemOpen
  \bibfield  {author} {\bibinfo {author} {\bibfnamefont {Masaki}\ \bibnamefont
  {Tezuka}}, \bibinfo {author} {\bibfnamefont {Onur}\ \bibnamefont {Oktay}},
  \bibinfo {author} {\bibfnamefont {Enrico}\ \bibnamefont {Rinaldi}}, \bibinfo
  {author} {\bibfnamefont {Masanori}\ \bibnamefont {Hanada}}, \ and\ \bibinfo
  {author} {\bibfnamefont {Franco}\ \bibnamefont {Nori}},\ }\bibfield  {title}
  {\enquote {\bibinfo {title} {Binary-coupling sparse sachdev-ye-kitaev model:
  An improved model of quantum chaos and holography},}\ }\href {\doibase
  10.1103/PhysRevB.107.L081103} {\bibfield  {journal} {\bibinfo  {journal}
  {Phys. Rev. B}\ }\textbf {\bibinfo {volume} {107}},\ \bibinfo {pages}
  {L081103} (\bibinfo {year} {2023})}\BibitemShut {NoStop}%
\bibitem [{\citenamefont {Kos}\ \emph {et~al.}(2018)\citenamefont {Kos},
  \citenamefont {Ljubotina},\ and\ \citenamefont {Prosen}}]{Kos_PRX_2018}%
  \BibitemOpen
  \bibfield  {author} {\bibinfo {author} {\bibfnamefont {Pavel}\ \bibnamefont
  {Kos}}, \bibinfo {author} {\bibfnamefont {Marko}\ \bibnamefont {Ljubotina}},
  \ and\ \bibinfo {author} {\bibfnamefont {Toma\ifmmode
  \check{z}\else~\v{z}\fi{}}\ \bibnamefont {Prosen}},\ }\bibfield  {title}
  {\enquote {\bibinfo {title} {Many-body quantum chaos: Analytic connection to
  random matrix theory},}\ }\href {\doibase 10.1103/PhysRevX.8.021062}
  {\bibfield  {journal} {\bibinfo  {journal} {Phys. Rev. X}\ }\textbf {\bibinfo
  {volume} {8}},\ \bibinfo {pages} {021062} (\bibinfo {year}
  {2018})}\BibitemShut {NoStop}%
\bibitem [{\citenamefont {Balabanov}\ and\ \citenamefont
  {Johannesson}(2017)}]{Bala_prb_2017}%
  \BibitemOpen
  \bibfield  {author} {\bibinfo {author} {\bibfnamefont {Oleksandr}\
  \bibnamefont {Balabanov}}\ and\ \bibinfo {author} {\bibfnamefont {Henrik}\
  \bibnamefont {Johannesson}},\ }\bibfield  {title} {\enquote {\bibinfo {title}
  {Robustness of symmetry-protected topological states against time-periodic
  perturbations},}\ }\href {\doibase 10.1103/PhysRevB.96.035149} {\bibfield
  {journal} {\bibinfo  {journal} {Phys. Rev. B}\ }\textbf {\bibinfo {volume}
  {96}},\ \bibinfo {pages} {035149} (\bibinfo {year} {2017})}\BibitemShut
  {NoStop}%
\bibitem [{\citenamefont {Liu}\ and\ \citenamefont {Guo}(2018)}]{Liu_pla_2018}%
  \BibitemOpen
  \bibfield  {author} {\bibinfo {author} {\bibfnamefont {Tong}\ \bibnamefont
  {Liu}}\ and\ \bibinfo {author} {\bibfnamefont {Hao}\ \bibnamefont {Guo}},\
  }\bibfield  {title} {\enquote {\bibinfo {title} {Topological phase transition
  in the quasiperiodic disordered su–schriffer–heeger chain},}\ }\href
  {\doibase https://doi.org/10.1016/j.physleta.2018.09.023} {\bibfield
  {journal} {\bibinfo  {journal} {Physics Letters A}\ }\textbf {\bibinfo
  {volume} {382}},\ \bibinfo {pages} {3287--3292} (\bibinfo {year}
  {2018})}\BibitemShut {NoStop}%
\bibitem [{\citenamefont {Meier}\ \emph {et~al.}(2018)\citenamefont {Meier},
  \citenamefont {An}, \citenamefont {Dauphin}, \citenamefont {Maffei},
  \citenamefont {Massignan}, \citenamefont {Hughes},\ and\ \citenamefont
  {Gadway}}]{Eric_science_2018}%
  \BibitemOpen
  \bibfield  {author} {\bibinfo {author} {\bibfnamefont {Eric~J.}\ \bibnamefont
  {Meier}}, \bibinfo {author} {\bibfnamefont {Fangzhao~Alex}\ \bibnamefont
  {An}}, \bibinfo {author} {\bibfnamefont {Alexandre}\ \bibnamefont {Dauphin}},
  \bibinfo {author} {\bibfnamefont {Maria}\ \bibnamefont {Maffei}}, \bibinfo
  {author} {\bibfnamefont {Pietro}\ \bibnamefont {Massignan}}, \bibinfo
  {author} {\bibfnamefont {Taylor~L.}\ \bibnamefont {Hughes}}, \ and\ \bibinfo
  {author} {\bibfnamefont {Bryce}\ \bibnamefont {Gadway}},\ }\bibfield  {title}
  {\enquote {\bibinfo {title} {Observation of the topological anderson
  insulator in disordered atomic wires},}\ }\href {\doibase
  10.1126/science.aat3406} {\bibfield  {journal} {\bibinfo  {journal}
  {Science}\ }\textbf {\bibinfo {volume} {362}},\ \bibinfo {pages} {929--933}
  (\bibinfo {year} {2018})}\BibitemShut {NoStop}%
\bibitem [{\citenamefont {Longhi}()}]{ste_opt_2020}%
  \BibitemOpen
  \bibfield  {author} {\bibinfo {author} {\bibfnamefont {Stefano}\ \bibnamefont
  {Longhi}},\ }\bibfield  {title} {\enquote {\bibinfo {title} {Topological
  anderson phase in quasi-periodic waveguide lattices},}\ }\href {\doibase Opt.
  Lett. 45, 4036-4039 (2020)} {\ Opt. Lett. 45, 4036-4039 (2020)}\BibitemShut
  {NoStop}%
\bibitem [{\citenamefont {Shi}\ \emph {et~al.}(2021)\citenamefont {Shi},
  \citenamefont {Kiorpelidis}, \citenamefont {Chaunsali}, \citenamefont
  {Achilleos}, \citenamefont {Theocharis},\ and\ \citenamefont
  {Yang}}]{Shi_prr_2021}%
  \BibitemOpen
  \bibfield  {author} {\bibinfo {author} {\bibfnamefont {Xiaotian}\
  \bibnamefont {Shi}}, \bibinfo {author} {\bibfnamefont {Ioannis}\ \bibnamefont
  {Kiorpelidis}}, \bibinfo {author} {\bibfnamefont {Rajesh}\ \bibnamefont
  {Chaunsali}}, \bibinfo {author} {\bibfnamefont {Vassos}\ \bibnamefont
  {Achilleos}}, \bibinfo {author} {\bibfnamefont {Georgios}\ \bibnamefont
  {Theocharis}}, \ and\ \bibinfo {author} {\bibfnamefont {Jinkyu}\ \bibnamefont
  {Yang}},\ }\bibfield  {title} {\enquote {\bibinfo {title} {Disorder-induced
  topological phase transition in a one-dimensional mechanical system},}\
  }\href {\doibase 10.1103/PhysRevResearch.3.033012} {\bibfield  {journal}
  {\bibinfo  {journal} {Phys. Rev. Res.}\ }\textbf {\bibinfo {volume} {3}},\
  \bibinfo {pages} {033012} (\bibinfo {year} {2021})}\BibitemShut {NoStop}%
\bibitem [{\citenamefont {Zuo}\ and\ \citenamefont
  {Kang}(2022)}]{Zuo_pra_2022}%
  \BibitemOpen
  \bibfield  {author} {\bibinfo {author} {\bibfnamefont {Zheng-Wei}\
  \bibnamefont {Zuo}}\ and\ \bibinfo {author} {\bibfnamefont {Dawei}\
  \bibnamefont {Kang}},\ }\bibfield  {title} {\enquote {\bibinfo {title}
  {Reentrant localization transition in the su-schrieffer-heeger model with
  random-dimer disorder},}\ }\href {\doibase 10.1103/PhysRevA.106.013305}
  {\bibfield  {journal} {\bibinfo  {journal} {Phys. Rev. A}\ }\textbf {\bibinfo
  {volume} {106}},\ \bibinfo {pages} {013305} (\bibinfo {year}
  {2022})}\BibitemShut {NoStop}%
\bibitem [{\citenamefont {\ifmmode~\check{S}\else \v{S}\fi{}untajs}\ \emph
  {et~al.}(2020)\citenamefont {\ifmmode~\check{S}\else \v{S}\fi{}untajs},
  \citenamefont {Bon\ifmmode~\check{c}\else \v{c}\fi{}a}, \citenamefont
  {Prosen},\ and\ \citenamefont {Vidmar}}]{unfolding2020PRE}%
  \BibitemOpen
  \bibfield  {author} {\bibinfo {author} {\bibfnamefont {Jan}\ \bibnamefont
  {\ifmmode~\check{S}\else \v{S}\fi{}untajs}}, \bibinfo {author} {\bibfnamefont
  {Janez}\ \bibnamefont {Bon\ifmmode~\check{c}\else \v{c}\fi{}a}}, \bibinfo
  {author} {\bibfnamefont {Toma\ifmmode \check{z}\else~\v{z}\fi{}}\
  \bibnamefont {Prosen}}, \ and\ \bibinfo {author} {\bibfnamefont {Lev}\
  \bibnamefont {Vidmar}},\ }\bibfield  {title} {\enquote {\bibinfo {title}
  {Quantum chaos challenges many-body localization},}\ }\href {\doibase
  10.1103/PhysRevE.102.062144} {\bibfield  {journal} {\bibinfo  {journal}
  {Phys. Rev. E}\ }\textbf {\bibinfo {volume} {102}},\ \bibinfo {pages}
  {062144} (\bibinfo {year} {2020})}\BibitemShut {NoStop}%
\bibitem [{\citenamefont {Kohmoto}\ \emph {et~al.}(1981)\citenamefont
  {Kohmoto}, \citenamefont {den Nijs},\ and\ \citenamefont
  {Kadanoff}}]{kohmotoPRB81}%
  \BibitemOpen
  \bibfield  {author} {\bibinfo {author} {\bibfnamefont {Mahito}\ \bibnamefont
  {Kohmoto}}, \bibinfo {author} {\bibfnamefont {Marcel}\ \bibnamefont {den
  Nijs}}, \ and\ \bibinfo {author} {\bibfnamefont {Leo~P.}\ \bibnamefont
  {Kadanoff}},\ }\bibfield  {title} {\enquote {\bibinfo {title} {Hamiltonian
  studies of the $d=2$ ashkin-teller model},}\ }\href {\doibase
  10.1103/PhysRevB.24.5229} {\bibfield  {journal} {\bibinfo  {journal} {Phys.
  Rev. B}\ }\textbf {\bibinfo {volume} {24}},\ \bibinfo {pages} {5229--5241}
  (\bibinfo {year} {1981})}\BibitemShut {NoStop}%
\bibitem [{\citenamefont {Prakash}\ \emph {et~al.}(2023)\citenamefont
  {Prakash}, \citenamefont {Fava},\ and\ \citenamefont
  {Parameswaran}}]{abhishodhPRL23}%
  \BibitemOpen
  \bibfield  {author} {\bibinfo {author} {\bibfnamefont {Abhishodh}\
  \bibnamefont {Prakash}}, \bibinfo {author} {\bibfnamefont {Michele}\
  \bibnamefont {Fava}}, \ and\ \bibinfo {author} {\bibfnamefont {S.~A.}\
  \bibnamefont {Parameswaran}},\ }\bibfield  {title} {\enquote {\bibinfo
  {title} {Multiversality and unnecessary criticality in one dimension},}\
  }\href {\doibase 10.1103/PhysRevLett.130.256401} {\bibfield  {journal}
  {\bibinfo  {journal} {Phys. Rev. Lett.}\ }\textbf {\bibinfo {volume} {130}},\
  \bibinfo {pages} {256401} (\bibinfo {year} {2023})}\BibitemShut {NoStop}%
\bibitem [{\citenamefont {Mondal}\ \emph {et~al.}(2023)\citenamefont {Mondal},
  \citenamefont {Agarwala}, \citenamefont {Mishra},\ and\ \citenamefont
  {Prakash}}]{mondal2023symmetryenriched}%
  \BibitemOpen
  \bibfield  {author} {\bibinfo {author} {\bibfnamefont {Suman}\ \bibnamefont
  {Mondal}}, \bibinfo {author} {\bibfnamefont {Adhip}\ \bibnamefont
  {Agarwala}}, \bibinfo {author} {\bibfnamefont {Tapan}\ \bibnamefont
  {Mishra}}, \ and\ \bibinfo {author} {\bibfnamefont {Abhishodh}\ \bibnamefont
  {Prakash}},\ }\href@noop {} {\enquote {\bibinfo {title} {Symmetry-enriched
  criticality in a coupled spin-ladder},}\ } (\bibinfo {year} {2023}),\ \Eprint
  {http://arxiv.org/abs/2309.04205} {arXiv:2309.04205 [cond-mat.str-el]}
  \BibitemShut {NoStop}%
\bibitem [{\citenamefont {Resta}(1998)}]{Resta_PRL_1998}%
  \BibitemOpen
  \bibfield  {author} {\bibinfo {author} {\bibfnamefont {Raffaele}\
  \bibnamefont {Resta}},\ }\bibfield  {title} {\enquote {\bibinfo {title}
  {Quantum-mechanical position operator in extended systems},}\ }\href
  {\doibase 10.1103/PhysRevLett.80.1800} {\bibfield  {journal} {\bibinfo
  {journal} {Phys. Rev. Lett.}\ }\textbf {\bibinfo {volume} {80}},\ \bibinfo
  {pages} {1800--1803} (\bibinfo {year} {1998})}\BibitemShut {NoStop}%
\end{thebibliography}
\end{document}